\documentclass[oldversion]{aa}
\usepackage{graphicx}
\usepackage{txfonts}
\usepackage{natbib}
\usepackage{multirow}
\usepackage{amssymb}
\usepackage{url}
\usepackage{rotating}

\def\2315{{RXC\,J2315.7-0222}}
\def\0216{{RXC\,J0216.7-4749}}

\def \xmm {\hbox{\it XMM-Newton}}

\begin{document} 

\title{Testing adiabatic contraction of dark matter in fossil group candidates}

\titlerunning{Adiabatic contraction in fossils}

\author{J. D\'emocl\`es \inst{1}, G.W. Pratt\inst{2}, D. Pierini\inst{3}, M. Arnaud\inst{2}, S. Zibetti\inst{4} and E. D'Onghia\inst{5}
        }

\authorrunning{J. D\'emocl\`es et al.}

\institute{
$^1$ IRFU/Service de Physique des Particules - CEA/DSM - CNRS, B\^{a}t. 141, CEA-Saclay, F-91191 Gif-sur-Yvette Cedex, France \\
$^2$ Laboratoire AIM, IRFU/Service d'Astrophysique - CEA/DSM - CNRS - Universit\'{e} Paris Diderot, Orme des Merisiers B\^{a}t. 709, CEA-Saclay, F-91191 Gif-sur-Yvette Cedex, France \\
$^3$ Max-Planck-Institut f\"ur extraterrestrische Physik, Giessenbachstra{\ss}e, 85748 Garching, Germany \\
$^{4}$ Max-Planck-Institut f\"ur Astronomie, K\"onigstuhl 17, D-69117 Heidelberg, Germany \\
${^5}$ Harvard Smithsonian Center for Astrophysics, 60 Garden Street, Cambridge MA, USA
}

\date{Received 26 February 2010 / Accepted 28 April 2010}
\offprints{J. Democles\\
           \email{jessica.democles@cea.fr}}

\abstract
{We present deep \xmm\ observations and ESO WFI optical imaging of two X-ray-selected fossil group candidates, \object{\0216} and \object{\2315}. Using the X-ray data, we derive total mass profiles under the hydrostatic equilibrium assumption. The central regions of \0216\ are found to be dominated by an X-ray bright AGN, and although we derive a mass profile, uncertainties are large and the constraints are significantly weakened due to the presence of the central source. The total mass profile of \2315\ is of high quality, being measured in fifteen bins from $[0.075 - 0.75]\,R_{500}$ and containing three data points interior to 30 kpc, allowing comprehensive investigation of its properties. We probe several mass models based on the standard NFW profile or on the  S\'ersic-like model recently suggested by high-resolution N-body simulations. We find that the addition of a stellar component due to the presence of the central galaxy is necessary for a good analytical model fit. In all mass profile models fitted, the mass concentration is not especially high compared to non-fossil systems.
 
In addition, the modification of the dark matter halo by adiabatic contraction slightly improves the fit. However, our result depends critically on the choice of IMF used to convert galaxy luminosity to mass, which leads to a degeneracy between the central slope of the dark matter profile and the normalisation of the stellar component. While we argue on the basis of the range of $M_*/L_R$ ratios that lower $M_*/L_R$ ratios are preferred on physical grounds and that adiabatic contraction has thus operated in this system, better theoretical and observational convergence on this problem is needed to make further progess.  }


\maketitle


\section{Introduction}                              
\label{sec:introduction}

The now well-established cold dark matter (CDM) paradigm lies at the heart of the fiducial scenario for the formation of structure in the Universe. Numerical simulations of hierarchical clustering in the currently-favoured $\Lambda$CDM cosmology make a number of observationally testable predictions. One example is the existence of a cusped, quasi-universal dark matter density profile that is characterised by the scale radius, $r_s$, at which the logarithmic slope of the profile is -2, and a dimensionless concentration parameter, $c$ \citep[e.g.,][]{nfw}. The latter parameter exhibits a distinct mass dependence related to the early formation of lower-mass haloes in the hierarchical context \citep[e.g.,][]{duf08}. 

Lying at the nodes of cosmic filaments, clusters and groups of galaxies are dark matter dominated objects whose properties are a sensitive test of these predictions. Indirect evidence for universality in the underlying dark matter distribution in galaxy groups and clusters was indicated by the similarity in X-ray surface brightness and temperature profiles observed with the {\it ROSAT\/} and {\it ASCA} satellites \citep[e.g.,][]{neu99,vik99,mar98}. However, more detailed, quantitative, information can be gleaned from examination of the total mass profiles of these systems obtained through the hydrostatic equilibrium equation. The vastly improved spatial resolution and throughput afforded by \xmm\ and {\it Chandra} have now allowed detailed investigation of the mass (and thus total density) profiles of moderately large samples of clusters and groups of galaxies \citep[e.g.,][and references therein]{pa02,pap05,vikh06,hum06,buo07,gasta07,sch07,sun09}. These and other works have confirmed the cusped, quasi-universal nature of the dark matter profile, with a variation in concentration roughly in line with predictions.

However, the presence of a large concentration
of baryons in the centre of the system can deepen
the potential well and modify the distribution of dark matter.
There has been extensive analytical and numerical work on the modification of the mass profile of a dark halo induced by the assembly of baryons in the inner regions.
Early results suggested that simple analytical prescriptions based
on the conservation of adiabatic invariants gave an
accurate description of the halo response. Following the early
work by \citet{bar84}, \citet{blu86}
devised a simple formula to link the dark mass profiles
before and after the assembly of a galaxy. Given the
initial, spherically-symmetric enclosed mass profiles of the dark
matter, $M^{i}_{\rm{dm}}$(r), and baryons, $M^{i}_b$(r), one may
derive the final dark mass profile, $M^f_{\rm{dm}}$(r),
once the final baryonic mass profile, $M^f_b$(r),
is specified. The model assumes that dark
matter particles move on circular orbits before and after the
contraction, and that their initial, $r_i$ , and final, $r_f$ ,
radii are related by the condition:

\begin{equation}
r_f[M_b^f (r_f)+M_{\rm{dm}}]=r_i [M_{\rm{dm}} + M^i_b(r_i)],
\end{equation}

\noindent where $M_{dm} = M_{\rm{dm}}(r_f) = M_{\rm{dm}}(r_i)$
is the dark mass enclosed by each dark matter particle. This modification is generally termed `adiabatic contraction' due to its parameterisation via adiabatic invariants \citep[e.g.,][]{egg62,blu86,gne04,duf10}. Further work by \citet{gne04} showed, based on hydrodynamic numerical simulations, that a modification to using the orbit-averaged radius described their results better than the standard circular orbit assumption. 

There are few observational tests of this prediction because of the difficulty of obtaining accurate mass profiles this deep into the core of a given object. Stellar dynamical studies suffer from the velocity dispersion anisotropy problem; weak lensing cannot measure the mass on these scales (less than 100 kpc), and strong lensing arcs are preferentially produced in dynamically disturbed systems \citep[e.g.,][]{bar96} which are unsuitable for the detection of the effect. Although some progress has recently been made on all of these fronts, \citep[e.g.,][]{san08}, in this context, X-ray observations offer several advantages conducive to investigation of the mass distribution in the central regions. In particular, the overall signal to noise of a good quality X-ray observation can be significant at many hundreds of sigma, and the density squared dependence of the emission results in a centrally-concentrated signal ideal for investigation of the core regions. The only assumption needed is that the intracluster medium (ICM) be in hydrostatic equilibrium. 

\citet{zap06} were the first to test for adiabatic contraction of the dark matter using X-ray observations. Applying several different analytical models (derived from numerical simulations) to the dark matter profile of the $\sim 3.5$ keV relaxed galaxy cluster \object{A2589}, they found no evidence for adiabatic contraction. \citet{gasta07} analysing the mass profiles of 16 galaxy groups, found that addition of adiabatic contraction did not improve the fit. Building on this result, \citet{hum06} examined the mass profiles of seven early type elliptical galaxies, again finding no strong evidence for adiabatic contraction.

Observational evidence points to the existence of a class of dark matter haloes that have the gravitating mass and hot gas content of groups, but which are dominated in the optical by a single large early-type galaxy \citep{pon94}. 
As noted by \citet{vik99}, such `fossil' systems may have been undisturbed for a very long time if they are the end result of merging within a normal compact group, and thus they may represent the ultimate examples of hydrostatic equilibrium\footnote{Note that some numerical simulations point to the fossil state being only a transient phase that is terminated by renewed infall from the local environment \citep{von08}.}. Furthermore, the exceptionally large concentration of baryons in the centre of the potential well makes them ideal systems with which to search for the effect of adiabatic contraction of the dark matter component.

In the present paper we build on our previous work on fossil systems \citep{zib09b} by investigating the mass profiles of two fossil group candidates, \0216\ and \2315, using X-ray and optical data. The central regions of \0216\ are dominated by emission from a bright AGN, and although we recover the total mass profile, the constraints are significantly weakened due to the presence of the AGN. For \2315, we are able to recover the mass profile to high precision. We find that the central regions are dominated by the stellar mass component, and using the mass profile of the central galaxy from optical data we investigate several parameterisations of the total mass distribution. We find that the best fitting mass parameterisation is obtained when we add a stellar component both in the case of a NFW and a \citet{nav04} profile, and that the addition of adiabatic contraction slightly improves the fit. However, the result is critically dependent on the assumed IMF used to convert stellar luminosity to mass. We discuss the impact of the assumed IMF on our results, and argue, based on the range of $M_*/L_R$ ratios, that adiabatic contraction has indeed operated in this system.

Unless otherwise noted, we adopt a $\Lambda$CDM cosmology with $H_0= 70$ km s$^{-1}$ Mpc$^{-1}$ (i.e., $h_{70} = 1$), $\Omega_M=0.3$ and $\Omega_\Lambda=0.7$. All uncertainties are quoted at the 68 per cent confidence level. In the following, $R_\delta$ is the radius corresponding to a density contrast of  $\delta$ times the critical density at the redshift of the system. 


\section{Optical data\label{sec:optical}}

\subsection{Observations}

Optical photometry of the two systems was carried out
using the Wide Field Imager \citep[WFI][]{baa99}
mounted at the Cassegrain focus of the MPG/ESO 2.2~m telescope
in La Silla, Chile.
The WFI is a focal reducer-type mosaic camera which consists of
$4 \times 2$ CCD chips, each with $2048 \times 4096$ pixels
and a field of view (FoV) of $8.12^{\prime} \times 16.25^{\prime}$
($0.238^{\prime \prime}$/pixel).
Chips are separated by gaps of $23.8^{\prime \prime}$
and $14.3^{\prime \prime}$ in the R.A. and Dec directions, respectively,
so the WFI FoV per exposure is equal to $34^{\prime} \times 33^{\prime}$,
with a filling factor of 95.9\%.
Imaging in the B ($\lambda_0 = 4562.52$ \AA) and R$_\mathrm{c}$
($\lambda_0 = 6517.25$ \AA, hereafter simply R) broad-band filters
was performed on September 19th, 2008 under photometric conditions
but with poor seeing ($\sim$1.7 and 1.6\arcsec\ FWHM in B- and R-band, respectively) owing to a strong wind. 
Twelve dithered observations gave a total exposure time
equal to 1080~s (B) and 720~s (R) and a resultant imaged region of
$30^{\prime} \times 30^{\prime}$.
The two fields of Landolt (1992) photometric standard stars SA\,92 and SA\,113
were imaged in the B, V ($\lambda_0 = 5395.62$ \AA), and R broad bands
in order to determine the photometric zero-point of the night.
Twilight-sky flats were taken in the evening and following morning
for each night of observations.

\begin{figure*}[]
\centering
\includegraphics[width=\columnwidth]{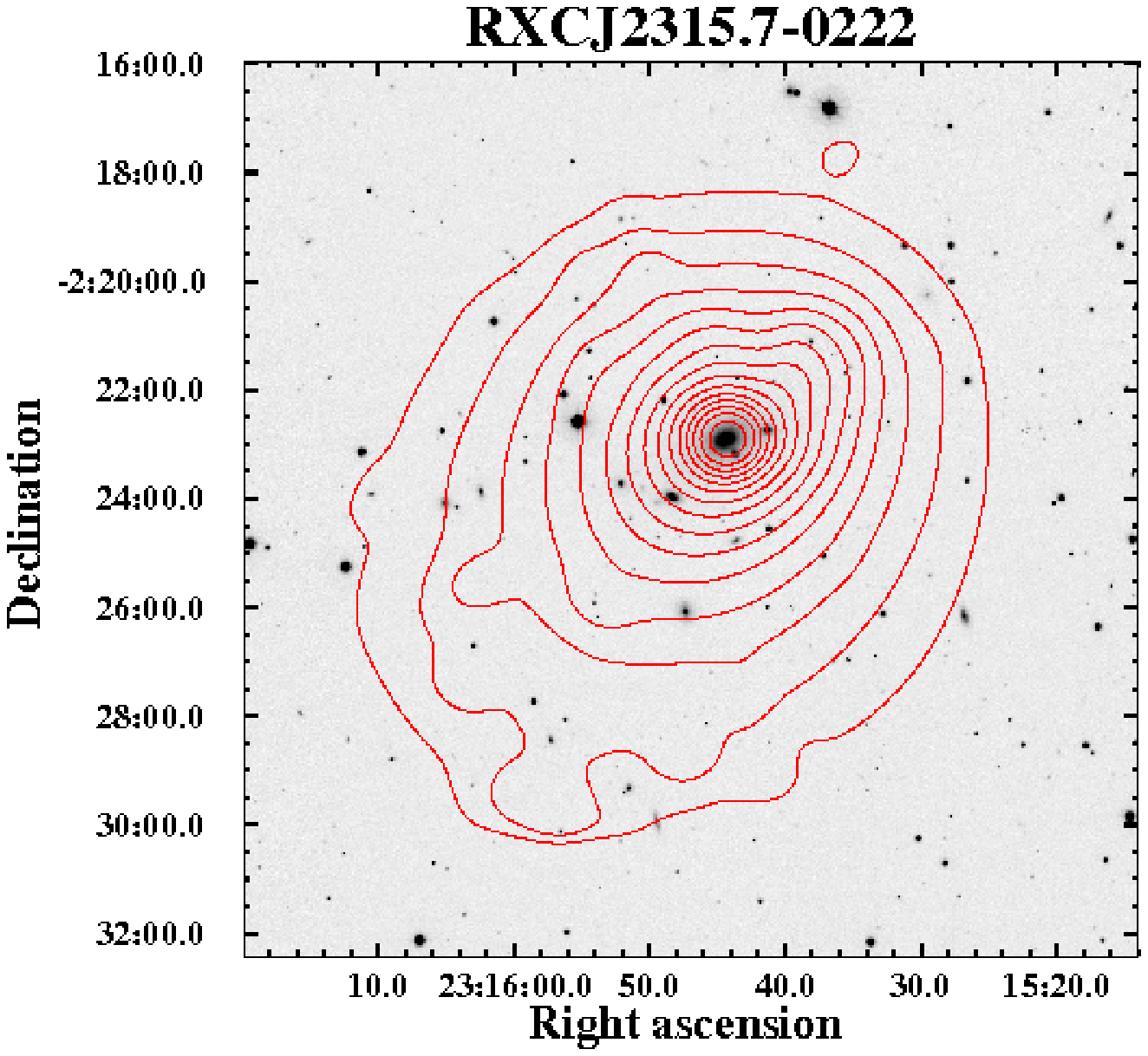} 
\hfill
\includegraphics[width=\columnwidth]{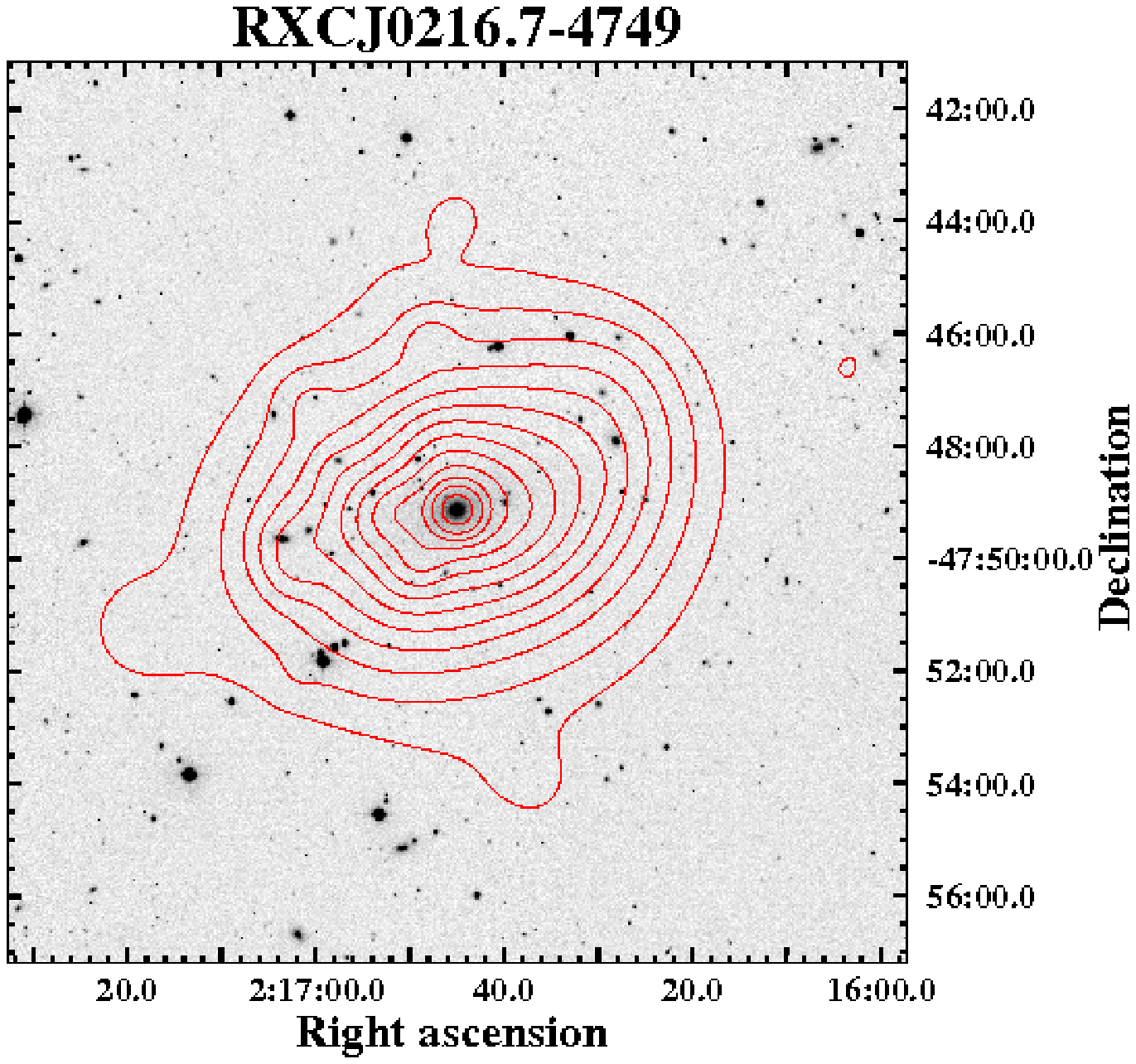} 
\label{Images}
\caption{\footnotesize $R$-band optical images of RXCJ2315.7-0222 (left) and RXCJ0216.7-4749 (right), obtained with the Wide Field Imager on the MPG/ESO 2.2m telescope at La Silla. The red contours start at $1\sigma$ above background and increase in steps of $\sqrt{2}$, and are derived from the wavelet-smoothed X-ray image. \label{fig:optim}}
\end{figure*}
\begin{figure}[]
\centering
\includegraphics[width=\columnwidth]{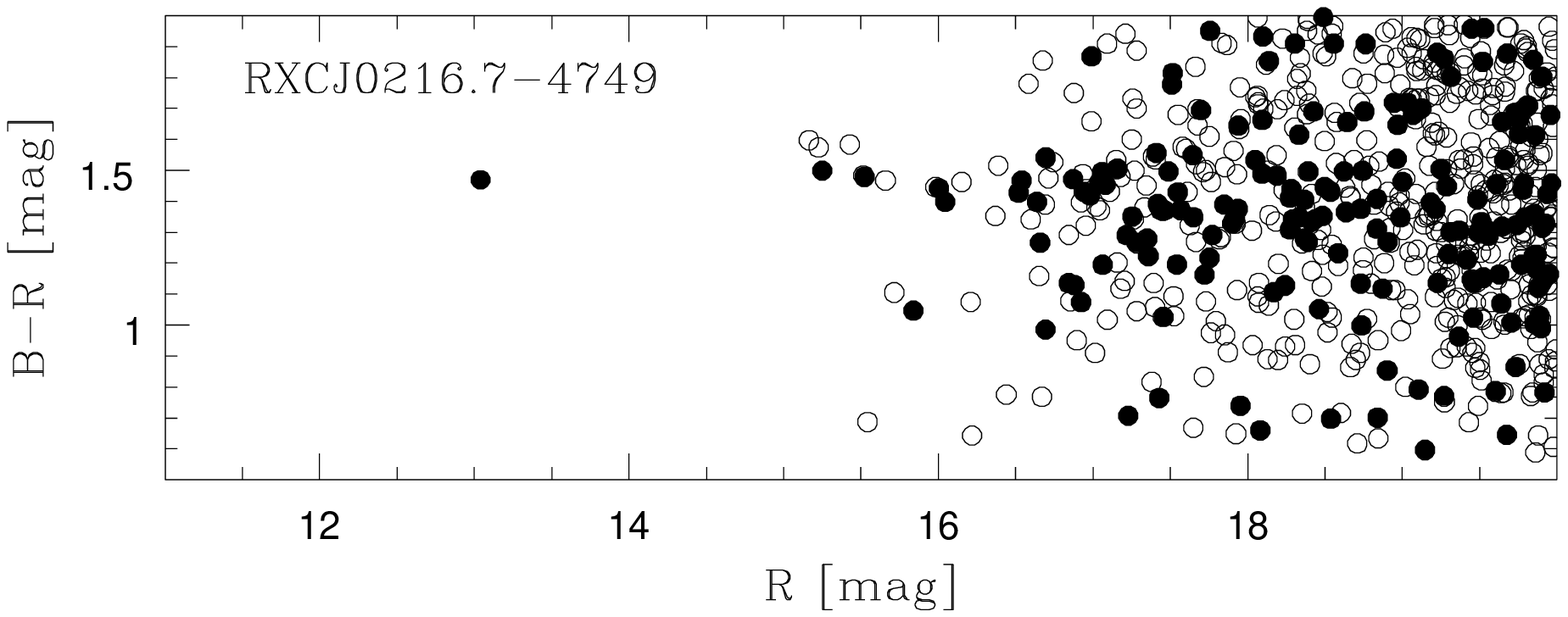} 
\hfill
\includegraphics[width=\columnwidth]{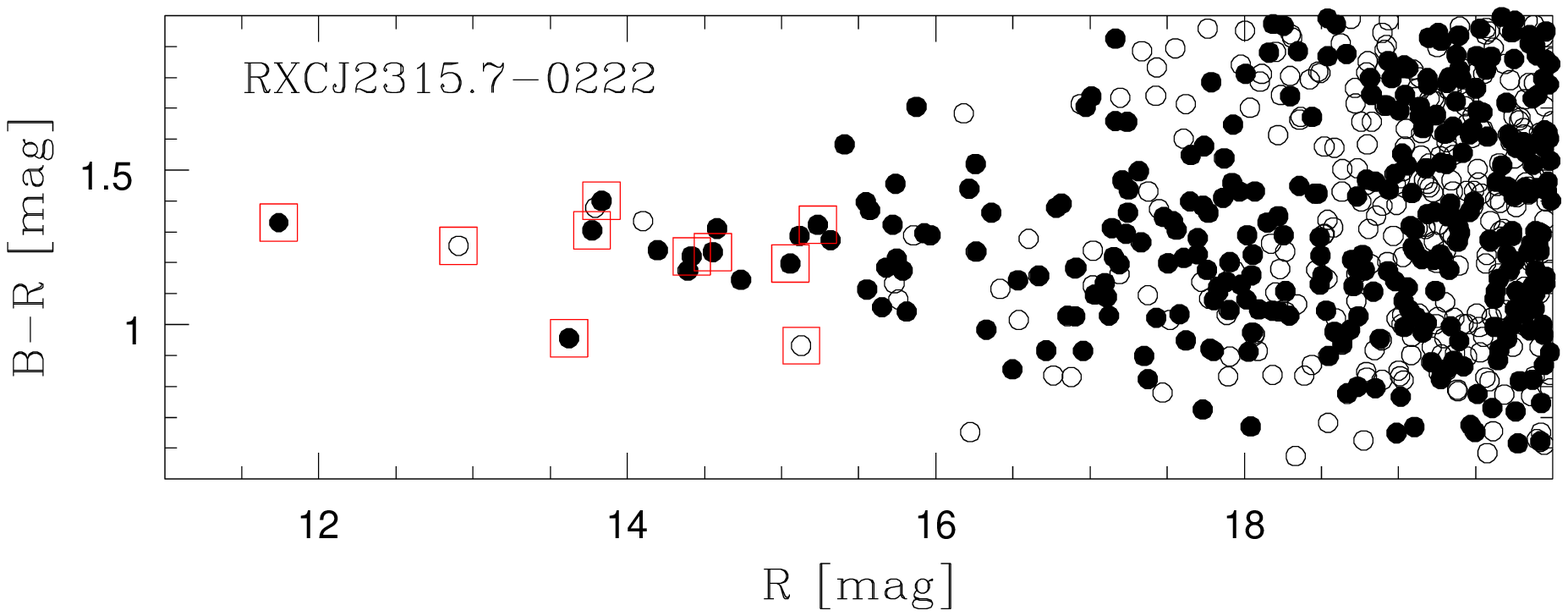} 
\caption{\footnotesize Colour-magnitude diagrams of \0216\ (top) and \2315\ (bottom). In both panels, filled circles denote objects within a projected radius of $0.5\, R_{200}$; open circles denote objects outside this radius. In the bottom panel, red squares denote spectroscopically confirmed members of \2315, as listed in NED.\label{fig:CMDs}}
\end{figure}


\subsection{Data reduction}

The WFI data were reduced using the data reduction system
developed for the ESO Imaging Survey \citep[EIS][]{ren97}
and its associated EIS/MVM image processing library version 1.0.1
({\it Alambic}). The data reduction steps are described in detail in previous publications \citep{arn01,mig07}, and are omitted here.

For images of standards, source detection and photometry
were based on SExtractor \citep{ber96}.
Magnitudes were calibrated to the Johnson-Cousins system
using \citet{lan92} standard stars
whose magnitudes were obtained using a 15~arcsec-wide circular aperture,
which proved to be adequate by monitoring the growth curve
of all the measured stars. Photometric solutions with minimum scatter (the uncertainty is equal to 0.03
and 0.05 mag for the zero points in the B and R passbands, respectively)
were obtained through a two-parameter linear fit to 24 photometric points
per passband, the extinction coefficient being set equal
to that listed in the ``definitive'' solution
obtained by the 2p2 Telescope Team.
In general, zeropoints and colour terms are consistent with those
obtained by the 2p2 Telescope Team or by the ESO DEEP Public Survey (DPS) team
\citep{mig07}. The final optical images of the two objects are shown in Figure~\ref{fig:optim}.

For the science frames, source detection and photometry were based on
SExtractor as well. Galaxies can be robustly separated from stars down to $R=20$ from the behaviour of the flux radius (i.e., the radius containing 50\% of the light)
  and the stellarity index as a function of magnitude.
  Visual inspection is used to test the selection criteria.
  At fainter magnitudes, a conservative separation can be obtained
  only by assuming a threshold value of 0.98 for the stellarity index,
  below which all selected, non-saturated/non-truncated objects
  are classified as bona-fide galaxies. For the brightest and largest
  galaxies within the value of $R_{200}$ of either group
  (likely member galaxies of RXCJ\,2315.7$-$0222) the Kron
\citep{kro80} magnitudes obtained through SExtractor were replaced
by the total magnitudes obtained by analytically extrapolating to
infinity the azimuthally averaged surface brightness (SB)
profiles. These are measured by means of elliptical isophote fitting
through the IRAF task \texttt{ellipse} as detailed in \citet{gav01}.  Spurious overlapping
stars/galaxies were conservatively masked out before elliptical
isophotes are fitted to the galaxy of interest.  Each SB profile was
either fitted using a single `de Vaucouleurs' $r^{1/4}$-law
component \citep{dev48} or, alternatively,
decomposed into an exponential law plus an inner `de Vaucouleurs'
$r^{1/4}$-law.  Fits were performed assigning the same weight to each
data point (logarithmically spaced) in order to properly weight the
outer parts of the profiles, which would be highly under-weighted in
a scheme based on the signal-to-noise ratio (S/N).  Furthermore,
points at semi-major axis smaller than the seeing FWHM were excluded.
Apparent surface brightness profiles and, thus, total magnitudes were
corrected for atmospheric extinction and galactic extinction \citep{sch98}. Figure~\ref{fig:CMDs} shows the colour-magnitude diagrams of the two systems, which are discussed further in Sect.~\ref{sec:fstatus}.

For \object{NGC\,7556}, the brightest cluster galaxy (BCG) of \2315, the corrected R-band surface brightness profile was integrated along the radial distance to give a
projected R-band luminosity profile, adopting a luminosity distance of
$104~h_{70}^{-1}~\mathrm{Mpc}$.  This luminosity profile was then
deprojected through the algorithm developed by
\citet{mag99} for an axysimmetric galaxy, with the kind help of
J. Thomas.  This algorithm finds the full range of smooth axisymmetric
(luminosity) density distributions consistent with a given surface
brightness distribution and inclination angle. The best solution
corresponds to the (luminosity) density that maximizes a penalized
log-likelihood function \citep[see][]{mag99}. 



\section{X-ray data \label{sec:Xray}}

\subsection{Preliminaries}

\0216\ and \2315\ were observed by \xmm\ in {\sc THIN} filter mode for 46 ks and 43 ks, respectively. Calibrated event files were produced using the standard {\tt chain}s in SASv8.0, and data were cleaned for soft proton flares, PATTERN selected and vignetting corrected as detailed in \citet{pratt07}. The resulting exposure times are 40 ks for \0216\ and 33 ks for \2315.

Point sources were detected on a broad band image via the SAS task {\tt ewavdetect}, with a detection threshold of $5\sigma$. After visual checking of the results, these sources were excluded from further analysis (excepting the central AGN in \0216\, which is discussed in more detail in the Appendices).


\subsubsection{Background subtraction}\label{sec:background}

Once flares are removed, the remaining background can be separated into two main components:

\begin{itemize}

\item the particle background, which is dominant at high energy, and 

\item the cosmic X-ray background (CXB), consisting of, at high energy, the unresolved AGN, and at low energy, emission from the Local Bubble and the halo of the Galaxy.

\end{itemize}

Using a custom stacked background assembled from observations taken with the filter wheel closed (FWC), recast to the same observation aspect as the source files, we adopt the following procedure for the background subtraction:

\begin{enumerate}
\item We use the count rates in a high energy band ([10-12] keV for EMOS, [12-14] keV for EPN) to normalise the FWC products to those of our observation.

\item For the surface brightness profiles, we subtract the normalised FWC data and then determine a region external to the group emission where the surface brightness profile is flat. From this area we determine the count rate due to the CXB, which we then subtract from the surface brightness profile.

\item For the spectral analysis, we extract a spectrum from the region external to the group emission, which we fit using a double unabsorbed thermal emission model for the galaxy and the local bubble, plus a powerlaw with fixed slope 1.4, absorbed with the Galactic column density in the direction of the group. This background spectrum, scaled by the ratio of the areas, is then added as an extra component in each annular spectrum of the temperature profile\footnote{Energy bands corresponding to fluorescence lines are ignored during the fit.}.

\end{enumerate}

\2315 fills the \xmm\ field of view, which complicates the background subtraction. We discuss the background subtraction for this object in Appendix~\ref{app:2315background}.


\subsection{Gas density distribution}

\subsubsection{Morphology}

Figure~\ref{Images} shows an \xmm/ESO WFI X-ray/optical overlay image for each system. Each fossil group exhibits morphologically regular emission with the X-ray peak centred on the brightest cluster galaxy, as expected if the systems have been dynamically quiescent for a considerable period of time. 


\begin{figure*}[]
\includegraphics[width=0.33\textwidth]{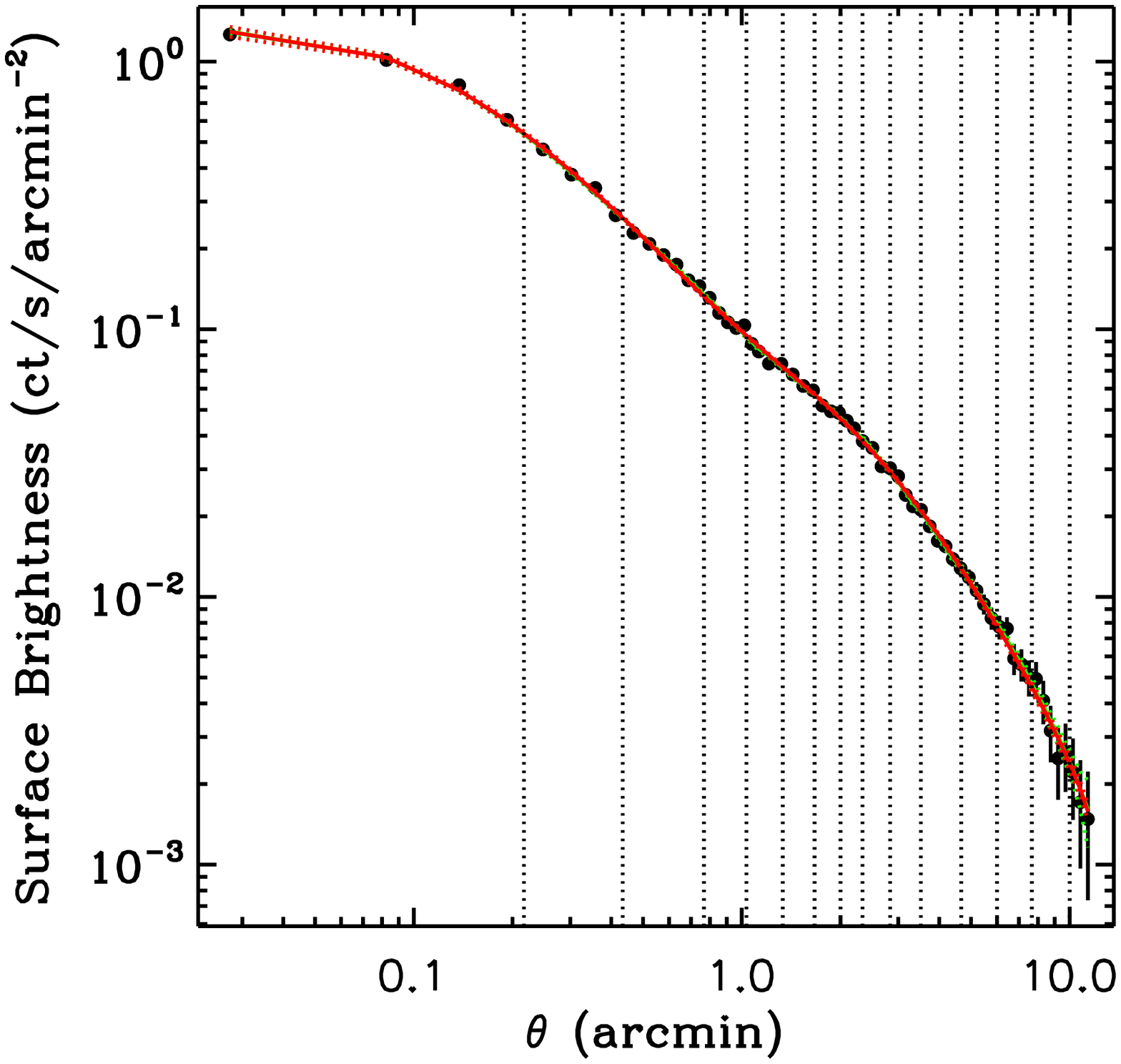} 
\hfill
\includegraphics[width=0.33\textwidth]{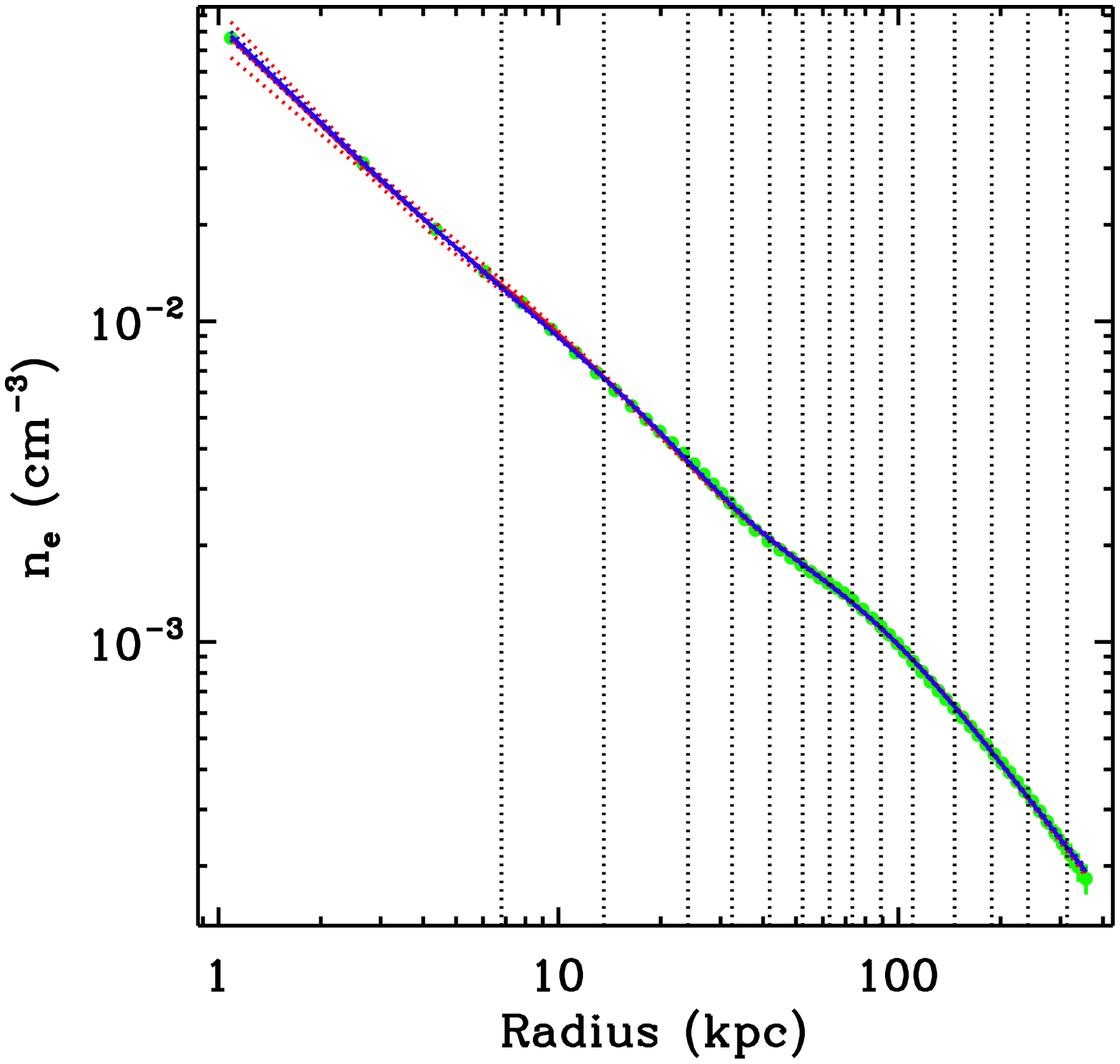} 
\hfill
\includegraphics[width=0.32\textwidth]{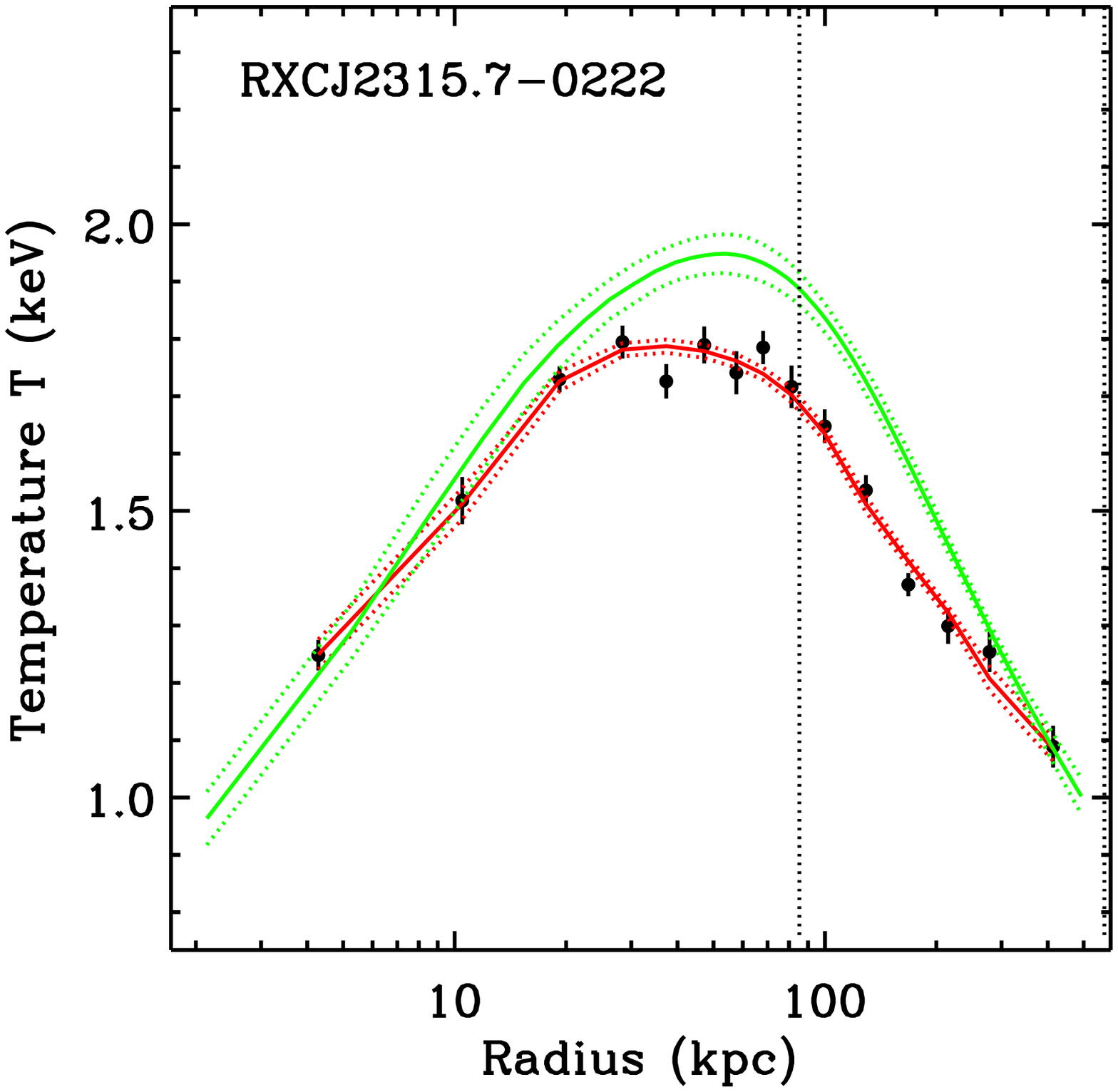} 
\caption{Radial profiles of \2315. {\it Left panel}:  Background subtracted, vignetting corrected surface brightness profile with best fitting projected analytical model (red) and projected deconvolved profile using the method of \citep{croston06} (green) overlaid.  Vertical dashed lines illustrate the extent of the annular bins from which spectra were accumulated. {\it Centre panel}: Corresponding density profiles. The solid blue line is the density profile derived from the best fitting AB model \citep{pa02} to the surface brightness profile. {\it Right panel}: Temperature profile. The black points with error bars depict the projected temperature profile. The solid green line is the  deprojected, PSF corrected 3D profile; dotted lines show the associated uncertainties. The solid red line is the reprojected 3D model.  \label{fig:2315prof}}
\end{figure*}

\begin{figure*}[]
\includegraphics[width=0.33\textwidth]{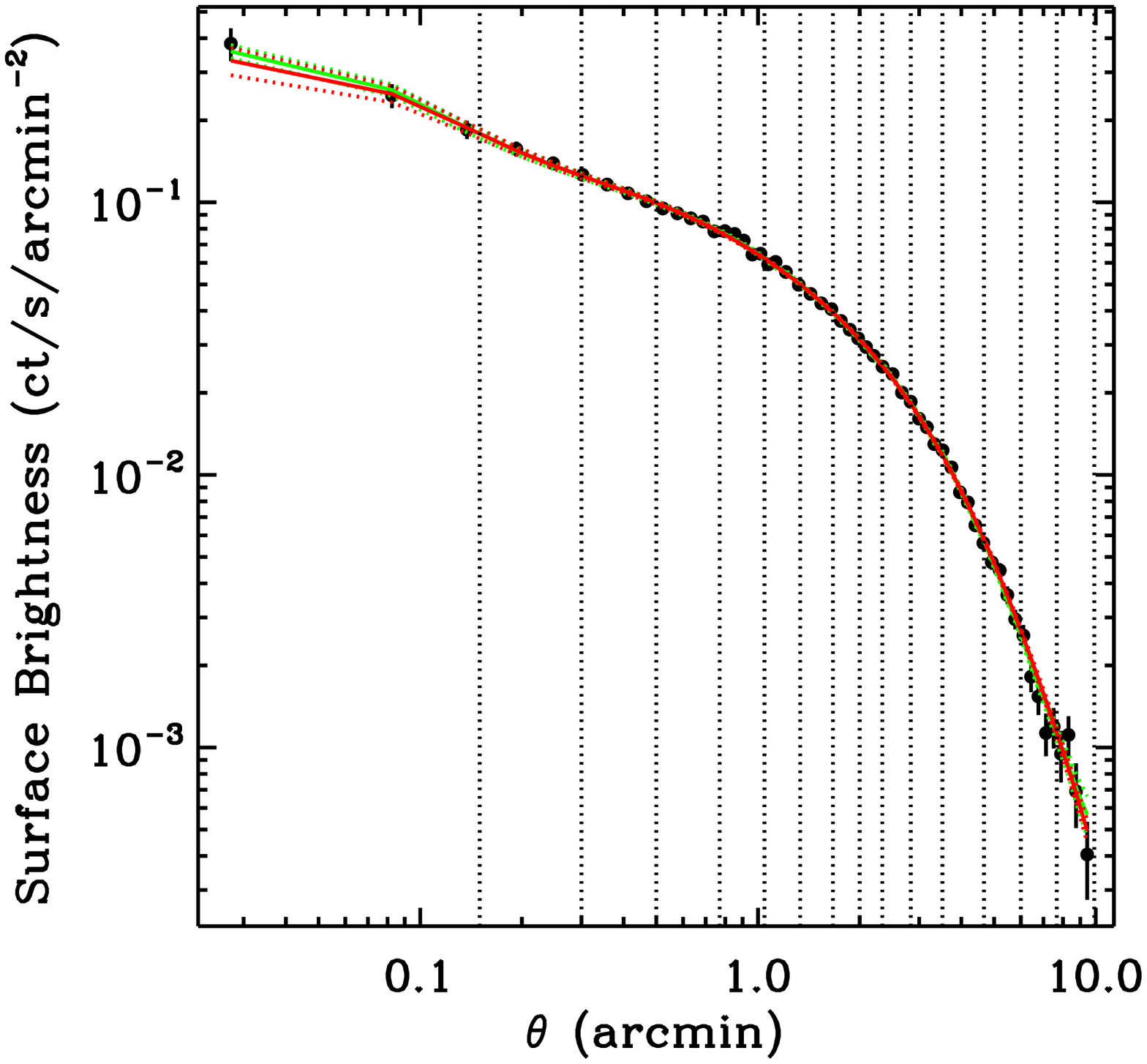} 
\hfill
\includegraphics[width=0.33\textwidth]{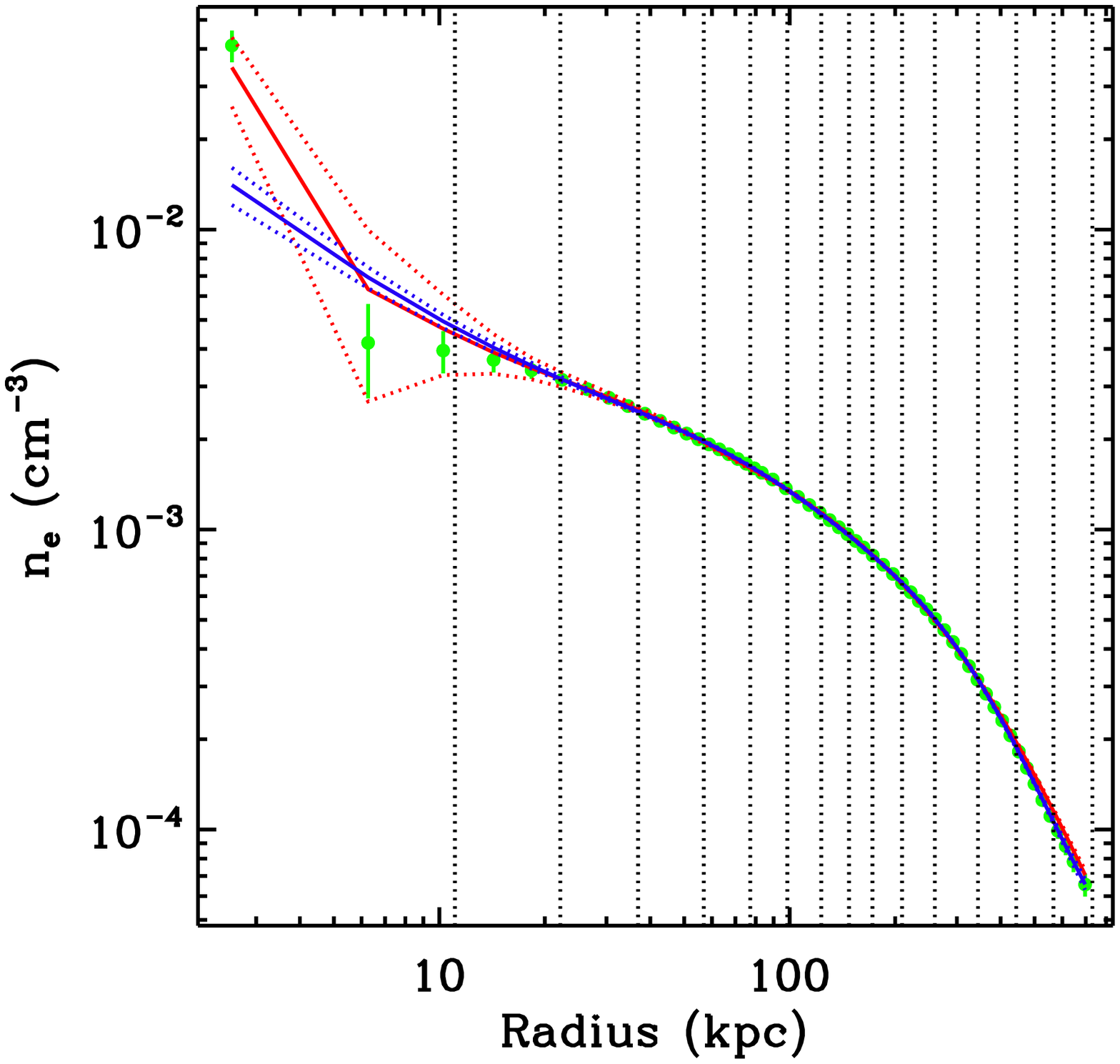} 
\hfill
\includegraphics[width=0.32\textwidth]{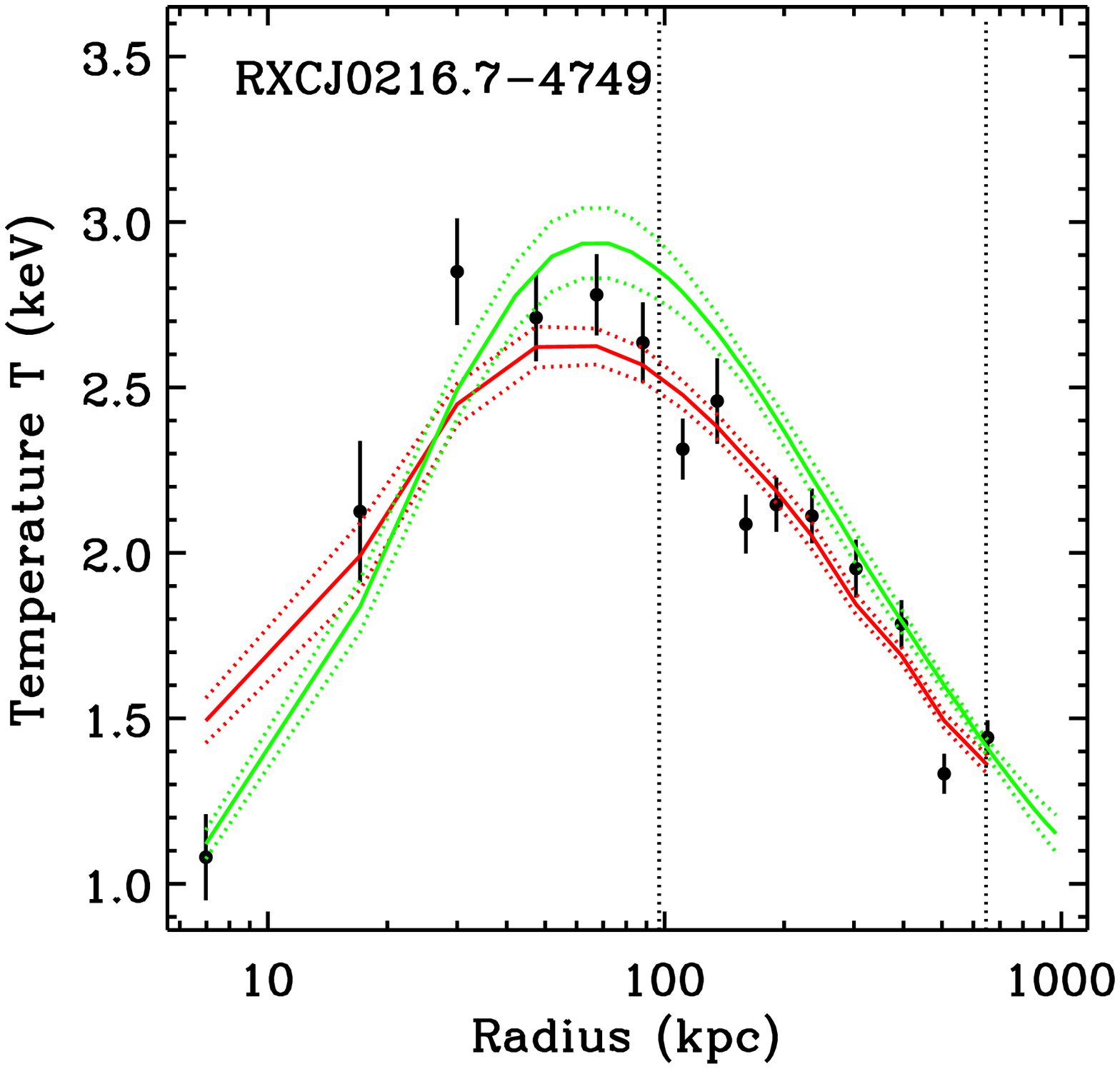} 
\caption{Radial profiles of \0216. Left to right: surface brightness profile, density profile, temperature profile. See Figure~\ref{fig:2315prof} for explanation of the line styles. \label{fig:0216prof}}
\end{figure*}

\subsubsection{Gas density profile}

The gas density profiles were derived as described in \citet{croston08}. Briefly, we extract the surface brightness profiles in $3\farcs3$ bins centred on the peak of the X-ray emission using photons from the [0.3-2] keV energy band. After background subtraction the profiles are rebinned to a significance of $3\sigma$ per bin. Deprojected, PSF-corrected emission measure profiles were produced using two methods: (i) the non-parametric method described in \citet{croston06} and (ii) by fitting a projected 3D parametric model \citep{vikh06}, convolved with the \xmm\ PSF, to the surface brightness profile. In both cases we use redistribution  matrices derived from the \citet{ghizzardi02} parameterisation of the \xmm\ PSF. Profiles were converted to gas density using a conversion factor derived in XSPEC from the global temperature measured in the [0.15-1]\,$R_{500}$ aperture. Finally, the temperature and abundance profiles were fitted with analytical models \citep[adapted from][]{asf01,vikh06} and used to calculate a correction factor for each density profile bin due to radial variations of these quantities. The surface brightness and gas density profiles of each group, derived using the two methods, are shown in the left and centre panels of Figures~\ref{fig:2315prof} and~\ref{fig:0216prof}. The parametric model is obviously smoother than the non-parametric deprojection, but the differences between the results from the two deconvolution methods are minimal. 

For \0216, the bright central AGN is evident as a point source in the X-ray surface brightness profile, which complicates deprojection and introduces significant uncertainty into the properties in the very central regions. For this object, we first fitted the surface brightness profile with an analytical model \citep{vikh06} plus a point source, with the point source normalisation  constrained from a spectral fit of the central region as detailed in Appendix~\ref{app:0216AGN}. We then subtract this contribution from the total surface brightness profile, adding 5 per cent systematic errors to the remainder, and deconvolve as detailed above.

\begin{table*}
\begin{minipage}{\textwidth}
\begin{center}
\caption{\footnotesize Basic data.\label{tab:bdata}}
\begin{tabular}{l c c c c c c c c}
\hline
\hline

Group & RA & Dec & $z$ & $L_{\rm X,bol}$\footnote{Bolometric ([0.001-100.] keV) luminosity calculated assuming $\Omega_M = 1$, $\Omega_\Lambda = 0$ and $h_0 = 0.5$.} & kT & $M_{500}$\footnote{Estimated from the $M_{500}-Y_X$ relation of \citet{app07}.} & $R_{500}$ & $R_{\rm det}$\footnote{Ratio of the weighted effective radius of the final bin of the temperature profile, $R_{\rm det}$, to $R_{500}$.} \\

&     &      &   &  (erg s$^{-1}$) & (keV) & $(\times 10^{13} M_{\odot})$ & (kpc) & $(R_{500})$ \\
\hline
\0216 & $02^{\rm h} 16^{\rm m} 45\fs0$ & $-47\degr 49\arcmin 09\farcs8$ &  0.064 & $3.2 \times 10^{43}$ & $2.05\pm0.05$ & $8.14\pm0.16$ & $645.7\pm5.0$ & 1.0 \\
\2315 & $23^{\rm h} 15^{\rm m} 44\fs4$ & $-02\degr 22\arcmin 54\farcs4$ & 0.026 & $2.1 \times 10^{43}$ & $1.68\pm0.03$ & $5.36\pm0.37$ & $568.7\pm13.1$ & 0.73 \\

\hline
\end{tabular}
\end{center}
\end{minipage}
\end{table*}


\subsection{Spectral analysis}

\subsubsection{Spectral fitting}

We extract spectra in circular annuli centered on the peak of X-ray emission. Annuli are defined so as to have a signal to noise ratio better than $30\sigma$ in the [0.3-2] keV energy range. After background subtraction they were binned to a minimum of 25 counts per bin. We then fitted an absorbed thermal bremsstrahlung model plus the background model described in Section~\ref{sec:background} to each spectrum using $\chi^2$ statistics. The background model has all parameters fixed and the normalisation of each component is scaled to the ratio of the extraction areas. Absorption is fixed to the Galactic value in the direction of the group. In the external regions of \0216\ the spectra are of insufficient quality to reliably constrain the abundances. In these annuli we freeze the abundance at $0.3\,Z_{\odot}$ and fit only the temperature and normalisation of the spectra. We fit in the [0.3-10] keV band excluding instrumental emission lines. The AGN component in \0216\ is modelled as described in Appendix~\ref{app:0216mprof}.


\subsubsection{Temperature profiles}

The projected temperature  profiles of the two groups are shown in  the right hand panels of Figures~\ref{fig:2315prof} and~\ref{fig:0216prof}. When plotted with a logarithmic radial axis, the temperature profiles for both of the groups exhibit the bell shape typical of cool core clusters \citep{vikh06,pratt07}. Deprojection and deconvolution of the temperature profiles is undertaken by fitting parametric 3D models \citep[adapted from][]{vikh06} to the projected profiles. These models are convolved with a response matrix that simultaneously takes into account projection and PSF redistribution, then projected and fitted to the observed temperature profile. In projecting the models, the weighting scheme introduced by \citet[][see also \citealt{mazz04}]{vikhw} is used to correct for the bias introduced by fitting isothermal models to multi-temperature plasma. Uncertainties are computed using a Monte Carlo procedure; these are subsequently corrected to take into account the fact that parametric models tend to over-constrain the 3D profile. Full details of the method will appear in a forthcoming paper.


\section{Reference mass, radius and fossil status}
\label{sec:fstatus}

We make a first estimate of $M_{500}$ of each system by iteration about the $M_{500}-Y_X$ relation of \citet{app07}, as described in \citet{kvn06}. $Y_X$ is the product of the gas mass inside $R_{500}$ and the temperature in the [0.15-0.75]\,$R_{500}$ region, a quantity which has been shown to be a robust, low scatter mass proxy in numerical simulations and observations \citep[e.g.,][]{app07,poole07}. The resulting temperatures, masses and radii are listed in Table~\ref{tab:bdata}. 

The empirical definition of a fossil group as established by \citet{jon03} requires a bolometric luminosity $L \geq 10^{42}\ h_{50}^{-2}$ erg s$^{-1}$ in X-rays, and a magnitude gap of $\Delta m_{12} \geq 2$ mag, where $\Delta m_{12}$ is the absolute total magnitude gap in $R$ band between the brightest and second brightest galaxies in the system within half the projected virial radius\footnote{\citeauthor{jon03} assume that the virial radius is equivalent to $R_{200}$.}. 

The spectroscopic bolometric luminosity of both systems is well above the $10^{42}\ h_{50}^{-2}$ erg s$^{-1}$ limit defined by (see Table~\ref{tab:bdata}). Regarding the magnitude gap, the colour-magnitude diagrams shown in Figure~\ref{fig:CMDs} clearly reveal a large magnitude gap between the brightest and second brightest galaxies in both systems. Assuming the values for $R_{500}$ listed in Table~\ref{tab:bdata}, and a ratio $R_{200}/R_{500} = 1.39$ \citep{pap05}, then $\Delta m_{12} = 2.21\pm0.03$ mag for \0216\, but is slightly less (at $\lesssim 2\sigma$) for \2315\ ($\Delta m_{12} = 1.88\pm0.03$)\footnote{Spectroscopic redshifts are available in NED for several group members in this case and, in particular, for the members brighter than $\mathrm{R} = 14$ which fall inside the region imaged with WFI.}. 

However, a system's fossil nature clearly depends critically on the definition of the virial radius. \citeauthor{jon03} used the global temperature in combination with a scaling relation derived from the numerical simulations in order to calculate their $R_{200}$; some of their global temperatures were estimated from the luminosity by assuming a different scaling relation. The $R_{500}$ (and thus $R_{200}$) measurements we use in this Section are derived from a scaling relation derived from X-ray hydrostatic mass measurements, and as shown below in Sect.~\ref{sec:massres}, $M_{500}$ varies by 20 per cent or so depending on the exact modelling of the mass profile, leading to changes on the order of an arcminute in the size of the projected $R_{200}$ aperture. Given the above considerations and the fact that the fossil criteria fixed by \citet{jon03} are somewhat arbitrary, while \2315\ does not {\it strictly} fulfil the criteria within the $1\sigma$ uncertainties on the photometry, we will continue to refer to it as a fossil system in the following. 

\begin{figure*}[]
\includegraphics[width=0.48\textwidth]{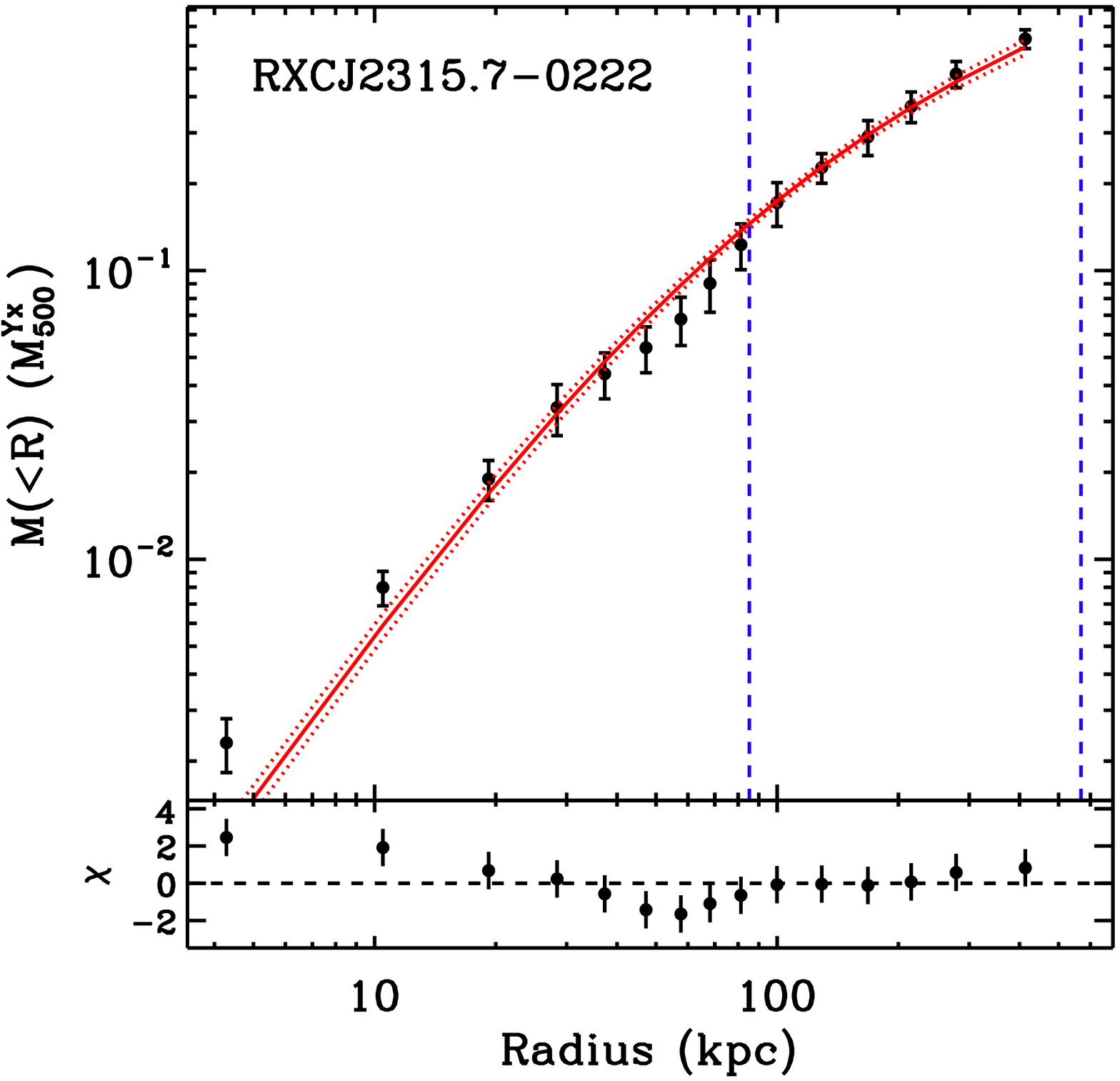} 
\hfill
\includegraphics[width=0.48\textwidth]{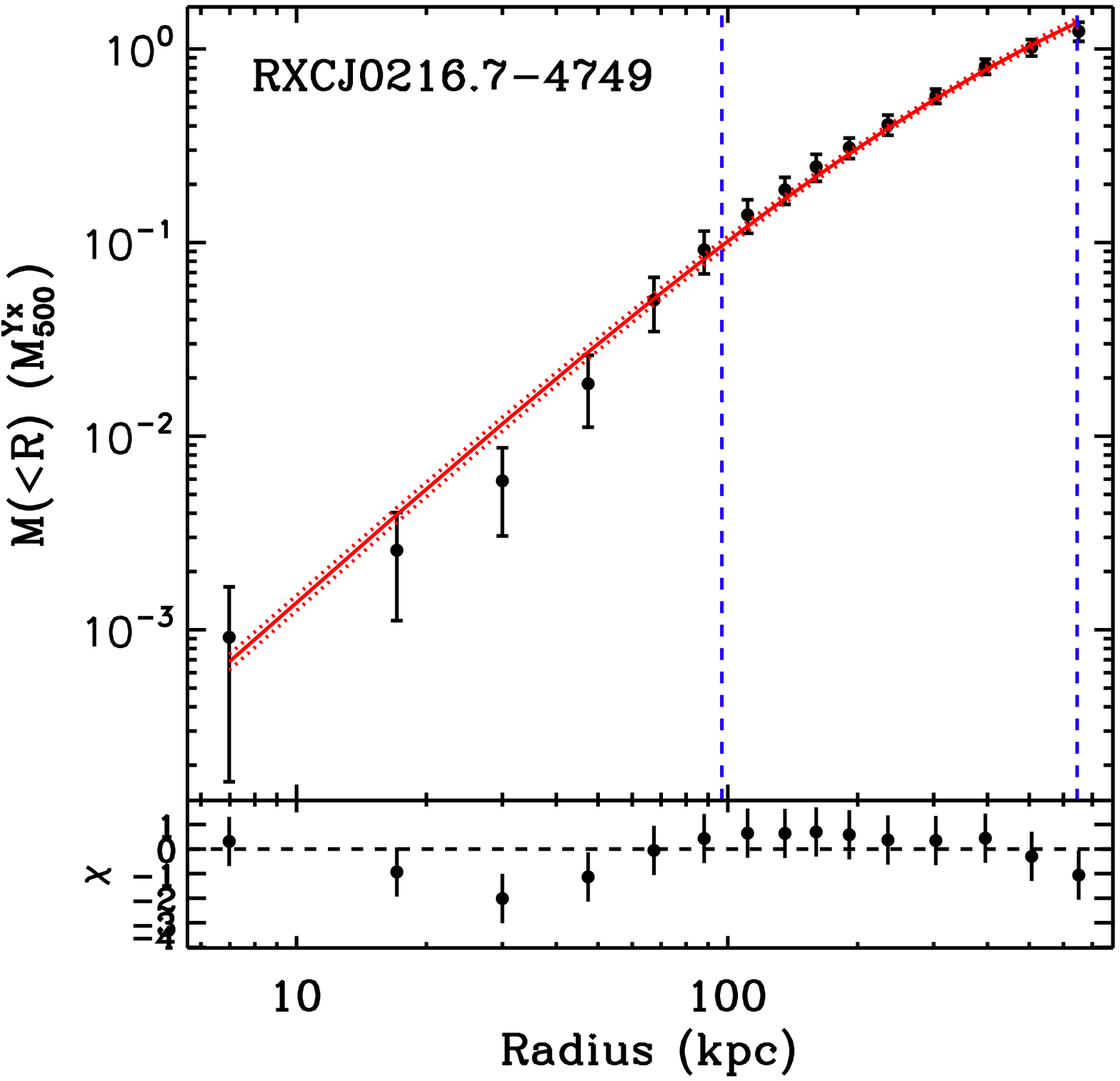} 
\caption{Hydrostatic total mass profiles for \2315\ (left) and \0216\ (right). Each profile is renormalised to the mass expected from the $M_{500}-Y_X$ relation of \citet{app07}. Vertical dashed lines show $r=0.15\, R_{500}$ and $r=R_{500}$. The solid red line is the best fitting NFW mass profile in each case.\label{fig:mprofs}}
\end{figure*}


\section{Mass profile modelling}


\subsection{Mass profile calculation}

We derive the total mass profile of each group from the density and temperature profiles (Figures~\ref{fig:2315prof} and~\ref{fig:0216prof}) using the hydrostatic equilibrium equation:

\begin{equation}
M(\leq R) = -\frac{kT(r)r}{G\mu m_p}\left[\frac{d\ln{n_e(r)}}{d\ln{r}}+ \frac{d\ln{T(r)}}{d\ln{r}}\right].
\end{equation}

\noindent The regularised gas density profiles exhibit structure that is amplified in the radial derivative. To overcome this, we fitted the 3D gas density profiles with the parametric model described by \citet{vikh06}. At each point corresponding to the effective radius of the deconvolved temperature profile, we then use the radial derivative given by the parametric function fit with uncertainties on $d \log{n_e}/ d \log{r}$ given by differentiation of the regularised density profile.

Uncertainties on each mass point are calculated using a Monte Carlo approach based on that of \citet{pa03}. A random temperature is generated at each radius at which the temperature profile is measured, and a cubic spline is used to compute the derivative. In the randomisation, we only keep profiles that are physical: they must increase monotonically with radius and the randomised temperature profiles must be convectively stable, i.e., $d \ln{T} / d\ln{n_e} < 2/3$. In total 1000 such Monte Carlo realisations were calculated; the error on the derivative is then the region containing 68 per cent of the realisations on each side. The resulting mass profiles are shown in Figure~\ref{fig:mprofs}. 


\subsection{Mass profile models\label{sec:massmods}}

We investigated various parameterisations of the total mass profile, as listed below.

\paragraph{NFW}: We first fitted these profiles with the integrated mass from an NFW profile \citep{nfw}, viz:

\begin{equation}
M(<r) = 4 \pi\, \rho_c(z)\, \delta_c\, r_s^3\, \left[ \ln{(1+r/r_s)} - \frac{r/r_s}{1+r/r_s} \right], \label{eqn:nfw}
\end{equation}

\noindent where $\rho_c(z)$ is the critical density of the universe at redshift $z$, $r_s$ is the scale radius where the logarithmic slope of the density profile reaches -2, and $\delta_c$ is a characteristic dimensionless density. This model has been shown to be an adequate fit to the mass profiles of many morphologically relaxed systems above $\sim2$ keV \citep[e.g.,][]{pa02,pap05,vikh06}. 

\paragraph{N04}: We also considered the universal profile derived from higher resolution simulations by \citet[][]{nav04}. The integrated mass of this profile is given by \citep[e.g.,][]{zap06}:

\begin{equation}
M(<r) = 4\pi\, \rho_{-2}\, r^3_{-2}\, \frac{1}{\alpha}\, \exp {\left( \frac{2}{\alpha} \right)} \left( \frac{\alpha}{2} \right)^{3/\alpha} \gamma \left( \frac{3}{\alpha} , \frac{2}{\alpha} x^{\alpha} \right)
\end{equation}

\noindent where $\gamma (\eta,\lambda) = \int_{0}^{\lambda} t^{\eta-1} e^{-t}\, dt$ is the lower incomplete gamma function, and $x=r/r_s$. As discussed below in Section~\ref{sec:massfits2315}, we fit both with $\alpha$ free and limited to lie within the $\pm1\sigma$ uncertainties found by \citet{nav04} (i.e., $0.14 \leq \alpha \leq 0.20$).

\paragraph{NFW+star and N04+star \label{sec:nstar}}: Previous investigations of group-scale haloes have suggested that below a certain mass scale (typically $\sim 2$ keV), the BCG begins to make a substantial contribution to the distribution of mass in the central regions \citep[e.g.,][]{gasta07,sun09}. As discussed in the Introduction, this contribution is potentially even more important in the case of a fossil group, where there is a considerable concentration of baryons in the centre of the potential well. 
We thus fitted the total mass profile with the NFW and N04 models detailed above with the addition of the mass from the central galaxy derived from our optical observations as described in Sect.~\ref{sec:optical}. The fits are initially undertaken with a fixed normalisation for the stellar component of $M_*/L_R = 1.84$, as obtained from the $B-R$ colour and the formulae given in \citet{zib09}, based on a \citet{chabrier03} IMF\footnote{By directly reverting to the model library of \citet{zib09} and excluding models with amounts of dust that are unrealistic for early-type galaxies, we obtain a median likelihood $M_*/L_R=1.41$ (thus 0.43 times lower than obtained with the above-mentioned fitting formula) with a $\pm 1~\sigma (3~\sigma)$ range of 1.03--1.80 (0.76--2.27). The adopted value of 1.84 should thus be regarded as an upper limit for a Chabrier IMF. Adopting a Salpeter (1955) IMF, all $M/L$ ratios would be scaled up by a factor of 1.75.}. 

In these fits, we first subtract the gas mass from the total gravitational mass profile to isolate the dark matter and stellar components. We then fit the resulting dark matter plus stellar mass with an NFW profile plus the stellar mass profile derived from our optical data (Section~\ref{sec:optical}). Since we are fitting the dark matter plus stellar mass profile, the concentration we derive is for these components only. In order to calculate the total mass concentration including the mass of the hot gas, we add the mass of the hot gas to that of the dark matter plus stars, and iterate \citep[see e.g.,][]{zap06}.

\paragraph{NFW+$k\,$star and N04+$k\,$star}: Adoption of different IMFs may result in mass estimates that differ by up to a factor 2, as shown by e.g.,  \citet[][their Figure~4]{bru03}. The exact form of the IMF is still a matter of debate (see Section~\ref{sec:IMF}) and therefore we also investigated the influence of freeing the normalisation of the stellar mass component. The total mass and corresponding NFW parameters are estimated by iteration as described above. We discuss the impact of the assumed IMF in detail in Sect.~\ref{sec:IMF}.

\paragraph{NFW*AC+$k\,$star and N04*AC+$k\,$star}: We also investigated the possibility of adiabatic contraction of the dark matter profile using the prescription described in \citet{gne04}, as implemented in the adiabatic contraction code {\tt contra}\footnote{{\tt http://www.astronomy.ohio-state.edu/$\sim$ognedin/contra/}.}. We choose this prescription among others \citep{blu86,aba09} for consistency with previous investigations \citep{zap06,hum06,gasta07}. In addition, {\tt contra} has been shown to perform well, in a statistical sense, in simulations of the dark matter response to galaxy formation for galaxies with total stellar masses at $z = 0$ in the range 2.5--$6 \times 10^{10}~h~\mathrm{M}_{\sun}$ having experienced little star formation over the last 8 Gyr \citep{tis09}. 

For input to the {\tt contra} code, we need the initial and final baryon profile and the initial dark matter profile. In this process only the stellar component is considered to affect the dark matter profile. We then proceed as follows: from the observed total mass profile we subtract the observed gas mass profile. We then fit with an NFW+$k\,$star or N04+$k\,$star model. Further subtracting the stellar mass component after this fit gives the initial dark matter profile used as input to {\tt contra}. The initial baryon (stellar) profile is assumed to follow the same form as the initial dark matter profile, with a normalisation given by:

\begin{equation}
M_{\rm stellar, init}(\leq r) = f_{\rm stellar} \times M_{\rm DM, init}( \leq r)$$
\end{equation}

\noindent where $f_{\rm stellar}$ is the stellar fraction at the maximum radius of the optical data. The final baryon profile is given by the observed stellar mass profile with appropriate normalisation. {\tt contra} is then run on the dark matter profile to compute the adiabatic contraction due to the stellar mass component. The resulting total mass profile is then compared to the data and the model iterated until convergence.

\begin{figure*}[]
\includegraphics[width=0.48\textwidth]{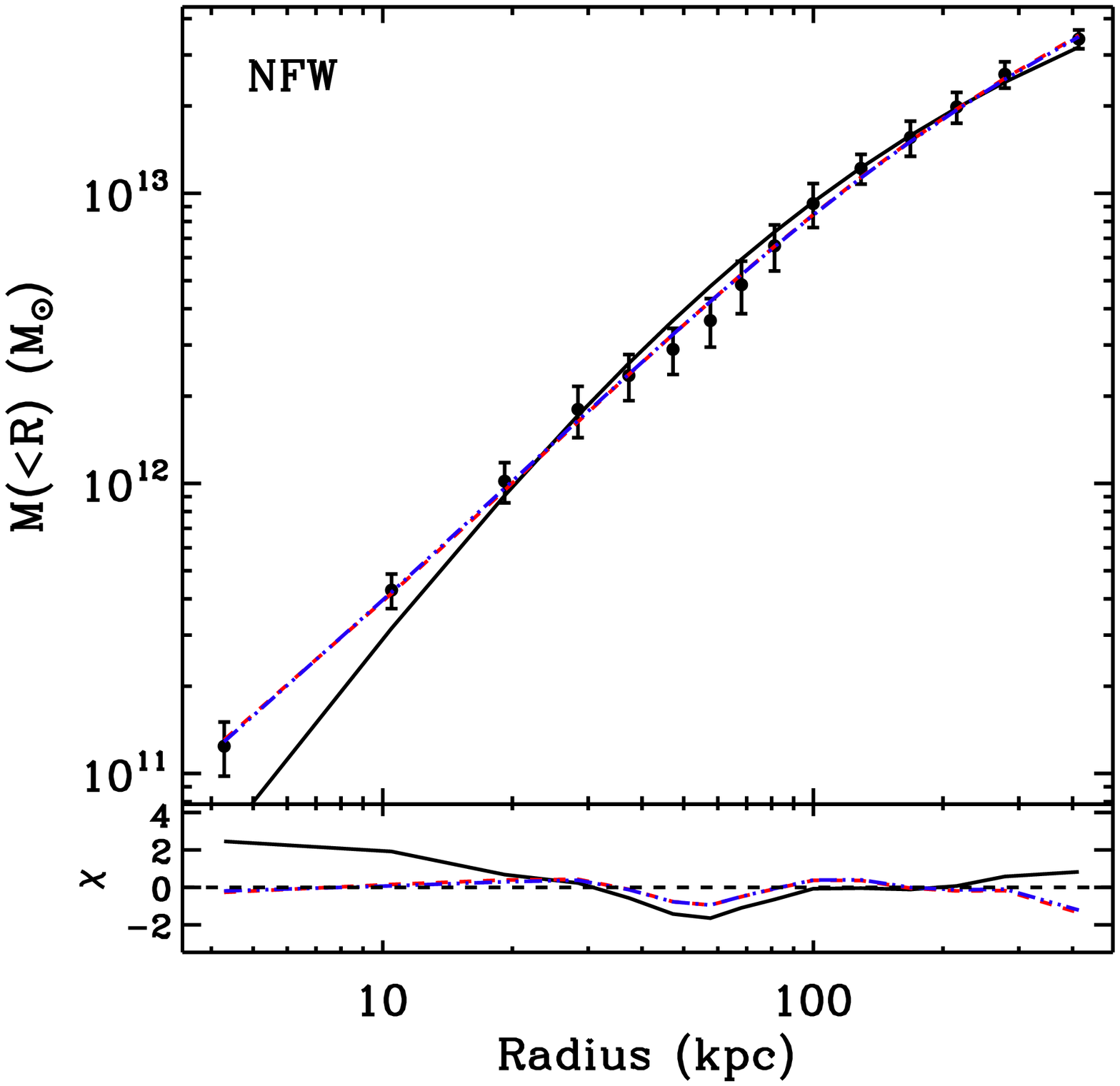} 
\hfill
\includegraphics[width=0.48\textwidth]{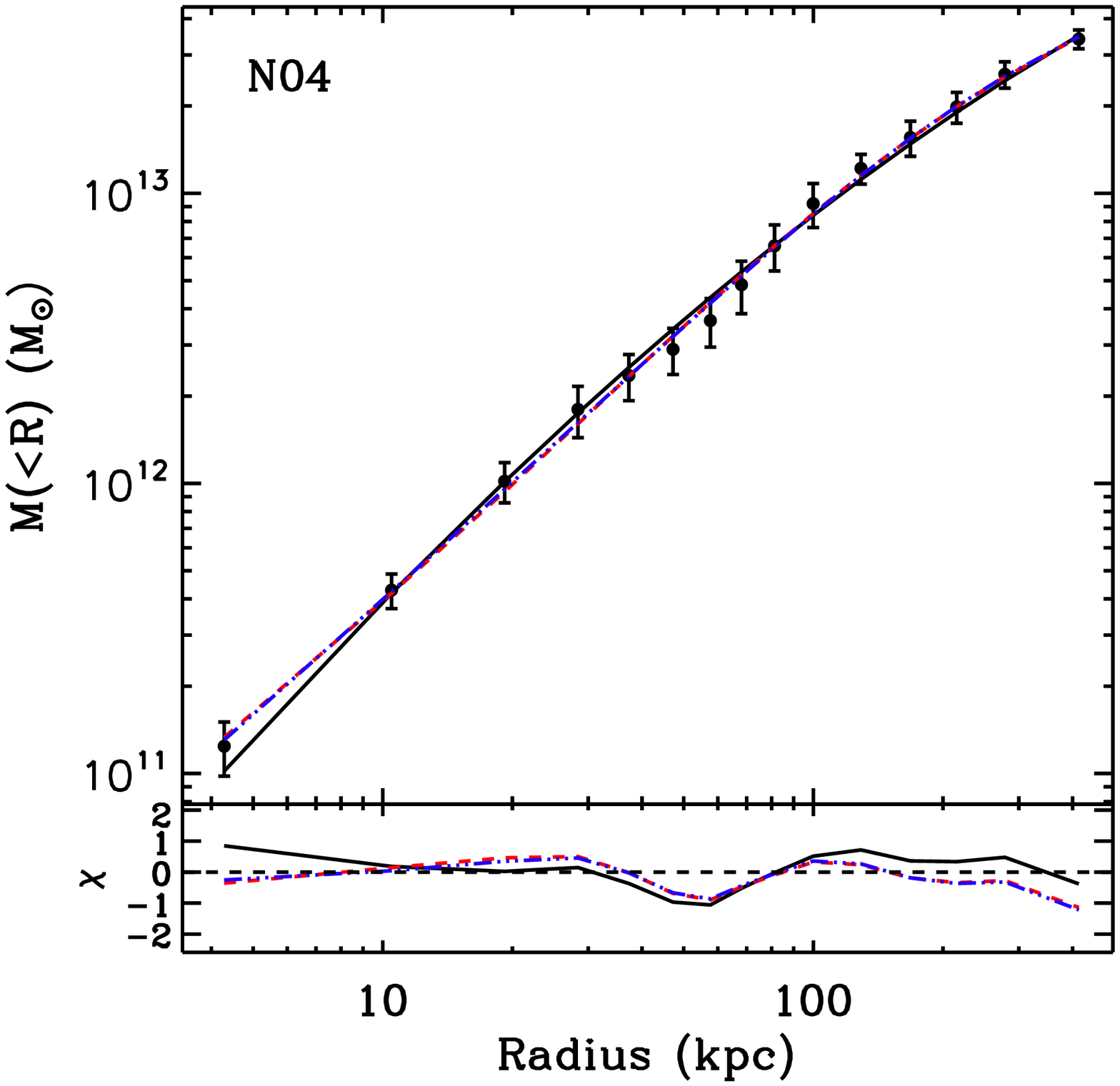} 
\caption{Additional modelling of the mass profile of \2315. {\it Left panel}: NFW model fitting. The solid line is the best fitting NFW model. The dashed line adds the stellar component with free normalisation (NFW$+k\,$star); the dot-dot-dot-dashed line adds adiabatic contraction to the dark matter component (NFW*AC$+k\,$star). {\it Right panel}: N04 fits, with same line styles. Corresponding best fitting parameters are listed in Table~\ref{tab:resfitmass}: the quality of the fit are roughly the equivalent with and without adiabatic contraction; the main differencies are in the normalisation of the stellar mass profile. \label{fig:2315mprofs}}
\end{figure*}

\begin{table*}
{\tiny 
\begin{minipage}{\textwidth}
\begin{center}
\caption{Results of fits to the mass profile of \2315.\label{tab:resfitmass}}
\begin{tabular}{l c c c c c r r r r}
\hline
\hline
Model & $c_{500}$ & $\alpha$ & $M_{500}$ $(10^{13}M_{\odot})$ & $R_{500}$ (kpc) & $f_{gas,500}$ & $M_{\star}/L_R$ & $M_{500}/M_{500}^{Yx}$ & $\chi^2/dof$ & dvi$_{\rm max}$ \\
\hline
\\
NFW                            & $8.01\pm0.82$ & \ldots & $3.58\pm 0.29$ & $497 \pm 14$ & $0.093\pm 0.013$ & \ldots & $0.67\pm0.07$ & $17.9/13$ & $1.08$ \\

NFW$+$star                     & $7.33\pm0.82$ & \ldots & $3.99\pm 0.33$ & $515\pm 14$ & $0.089\pm 0.013$ & 1.84 & $0.74\pm0.08$ & $7.9/13$ & $0.72$ \\

NFW$+\,k\,$star             & $5.68\pm0.93$ & \ldots & $4.33\pm 0.37$ & $530\pm 15$ & $0.086\pm 0.013$ & $4.40\pm 1.08$ & $0.81\pm0.09$ & $2.8/12$ & $0.14$ \\

NFW$^{\ast}$AC$+$star          & $6.01\pm0.59$ & \ldots & $3.99\pm 0.27$ & $515 \pm 12$ & $0.089\pm 0.012$ & $1.84$ & $0.74\pm0.07$ & $3.65/13$ & $0.15$ \\

NFW$^{\ast}$AC$+\, k\,$star & $5.61\pm0.12$ & \ldots & $4.17\pm 0.27$ & $523\pm 11$ & $0.087\pm 0.011$ & $1.81\pm0.33$ & $0.78\pm0.07$ & $2.67/12$ & $0.14$ \\

\\

$\alpha$ free \\
\cline{1-1\vspace{2mm}}
N04                            & $5.87\pm0.97$ & $0.108\pm0.016$ & $4.30\pm0.35$ & $528\pm14$ & $0.085\pm0.012$ & \ldots & $0.80\pm0.09$ & 4.63/12 & $0.22$ \\

N04+star                       & $5.53\pm1.37$ & $0.138\pm0.042$ & $4.42\pm0.31$ & $533\pm13$ & $0.085\pm0.011$ & 1.84 & $0.82\pm0.08$ & 3.78/12 & $0.15$ \\

N04+$k\,$star               & $5.30\pm1.04$ & $0.299\pm0.153$ & $4.03\pm0.31$ & $517\pm14$ & $0.088\pm0.012$ & $4.50\pm2.34$ & $0.75\pm0.08$ & 2.36/11 & $0.14$ \\

N04*AC+star                    & $5.07\pm0.95$ & $0.241\pm0.074$ & $4.20\pm0.30$ & $524\pm13$ & $0.086\pm0.012$ & 1.84 & $0.78\pm0.08$ & 2.53/12 & $0.14$ \\

N04*AC+$k\,$star            & $5.12\pm1.14$ & $0.326\pm0.182$ & $4.05\pm0.30$ & $518\pm13$ & $0.088\pm0.012$ & $2.09\pm1.18$ & $0.76\pm0.08$ & 2.09/11 & $0.13$ \\

\\

$0.14 \leq \alpha \leq 0.20$ limited \\
\cline{1-1\vspace{2mm}}

N04                            & $6.81\pm0.82$ & $0.142\pm0.005$ & $4.01\pm0.31$ & $516\pm13$ & $0.088\pm0.012$  & \ldots & $0.75 \pm0.08$ & 6.78/12 &  0.43\\
N04+star                       & $6.12\pm0.85$ & $0.154\pm0.021$ & $4.28\pm0.26$ & $527\pm11$ & $0.086\pm0.011$ & 1.84 & $0.80\pm0.08$ & 4.02/12 & 0.16\\
N04+$k\,$star              & $5.39\pm1.05$ & $0.189\pm 0.025$ & $4.30\pm0.28$ & $528\pm12$ & $0.086\pm0.011$ & $3.64\pm1.39$ & $0.80\pm0.08$ & 3.05/11 & 0.14\\
N04*AC+star                   & $4.83\pm0.65$ & $0.193\pm0.021$  & $4.43\pm0.25$ & $533\pm10$ & $0.084\pm0.010$ & 1.84 & $0.83\pm0.08$ & 3.29/12 & 0.13\\
N04*AC+$k\,$star           & $5.32\pm1.19$ & $0.191\pm0.024$ & $4.35\pm0.26$ & $530\pm11$ & $0.085\pm0.011$ & $1.51\pm0.71$ & $0.81\pm0.08$ & 3.01/11 & 0.14\\

\\

\hline
\end{tabular}
\end{center}
\end{minipage}
}
\end{table*}

As we will see below, the major contributions to $\chi^2$ come from the inner data points ($R \lesssim 30$ kpc), where the uncertainties on the temperature and density are smallest. Many of the models are formally acceptable, and it is not possible to distinguish between them purely in terms of $\chi^2$. Following the discussion in \citet{gasta07}, we thus also give the values of the maximum fractional deviation dvi$_{\rm max}$, which gives equal weight to all radial bins, as an additional discriminator of the goodness of fit. The dvi$_{\rm max}$ was extensively used by \citet{jin00} in an investigation of the goodness of fit of the NFW analytical form to dark matter haloes both in and out of equilibrium, where it was found that a dvi$_{\rm max} < 0.3$ represented a good fit of the NFW model. 


\subsection{Mass fitting results}
\label{sec:massres}

\subsubsection{\0216}

While the weighted effective radius \citep{lewis03} of the outermost temperature point lies at $\sim R_{500}$, the uncertainties on the mass profile of \0216\ are severely affected by the presence of the central AGN. As a result, a simple NFW model with $c_{500} = 2.22\pm0.25$ and $M_{500} = 1.31^{+0.12}_{-0.11} \times 10^{14}\,M_{\odot}$ is an acceptable fit ($\chi^2$/dof = 9.8/13) to the mass profile of this object once corrected for the AGN contribution. The AGN-corrected profile and best-fitting simple NFW model are shown in the right hand panel of Figure~\ref{fig:mprofs}. 

Partly the good fit is because of the large uncertainties in the inner regions which are the result of the correction for the central AGN. Appendix~\ref{app:0216mprof} compares the parameters of the NFW model fit to the mass profile of \0216\ in the cases of (i) no correction for the AGN contribution; (ii) excision of the central 18\arcsec, and (iii) modelling of the AGN contribution as described in Appendix~\ref{app:0216AGN}. The total mass varies by 30 per cent, with the lowest value derived for the uncorrected mass profile and the highest value derived for the mass profile corrected for the AGN contribution. The concentration parameter, unsurprisingly, varies in the opposite sense, being highest for the uncorrected profile and lowest for the corrected profile. 

The fact that we detect the mass profile directly to $\sim R_{500}$ gives us confidence that the result is not due to an extrapolation bias resulting from fitting the mass profile in a limited radial range. However, we do not pursue more complicated model fits to the mass profile of this system due to the limited quality of the data within $0.15\,R_{500}$.


\subsubsection{\2315}\label{sec:massfits2315}

In contrast, the mass profile data of \2315\ are of sufficiently good quality for a more detailed investigation: there are three well-constrained data points within 30 kpc, and the data extend to $\sim 0.8\,R_{500}$, allowing us to constrain both the mass distribution in the centre and the slope of the mass profile in the outer regions with good accuracy. Figure~\ref{fig:2315mprofs} shows the mass profile data and the various model fits which we will now discuss. 

The mass profile of \2315 is not well described by the simple NFW model: the reduced $\chi^2$ is rather high ($\chi^2$/dof = 17.9/13), and there is a systematic divergence of the data points from the model in the inner $\sim 100$ kpc (Figures~\ref{fig:mprofs} and~\ref{fig:2315mprofs}). The excess of mass in the inner regions with respect to the NFW profile is significant ($\sim 2.5\sigma$ for the inner point), suggesting that there is a substantial contribution in the central regions due to the stellar mass component. Indeed, the initial fit of an NFW+star profile with fixed normalisation of the stellar component results in a substantial improvement in $\chi^2$ relative to the simple NFW fit, and the central regions in particular are noticeably better fitted. The addition of the stellar mass component results in a decrease in the value of the concentration parameter, although this is not significant within the relatively large uncertainties. Further freeing the normalisation of the stellar mass component (NFW+$k\,$star) results in another improvement in $\chi^2$, which an f-test confirms is very highly significant (probability $5.3 \times 10^{-4}$). 

The N04 mass profile allows more centrally concentrated mass distributions than the original NFW parameterisation. Fitting this form to the total mass data with the $\alpha$ parameter left free results in a substantial improvement in the fit statistic over the simple NFW case ($\chi^2$/dof = 4.63/12), due primarily to a better fit of the inner three data points. However, the inferred value of $\alpha = 0.108\pm0.016$ is significantly lower than the mean value of $0.172\pm0.032$ found by \citet{nav04}, implying a density profile that is steeper in the centre and shallower at large radius than that inferred from CDM. Limiting $\alpha$ to lie within the $\pm1\sigma$ uncertainties found by \citet{nav04} (i.e., $0.14 \leq \alpha \leq 0.20$) results in a marginally degraded fit ($\chi^2$/dof = 6.78/12).

An N04+star model with $\alpha$ free and a fixed stellar normalisation results in a further improvement in $\chi^2$ over the simple N04 case and more interestingly, this fit yields a value of $\alpha$ that is in better agreement with the predictions from CDM. Further freeing the normalisation of the stellar component results in another improvement in $\chi^2$ but at the expense of a higher value of $\alpha$ and a higher $M_*/L_R$, although then neither parameter is well constrained. A similar trend is found for fits with $\alpha$ limited to lie within the $\pm1\sigma$ uncertainties found by \citet{nav04}.

Finally, we investigated the case where adiabatic contraction of the dark matter is taken into account. In both the NFW and N04 cases, adiabatic contraction of the dark matter profile improves the fit slightly relative to that without. Formally the best fitting model is an NFW profile with free stellar normalisation and adiabatic contraction applied to the dark matter, but the equivalent N04 profile is statistically almost as good a fit. We discuss the implications of this result, and in particular its dependence on the assumed IMF, in the next Section.


\section{Discussion}


\subsection{Total mass profile of \2315}

\subsubsection{Introduction}

In Section~\ref{sec:massfits2315} above we showed that the best fitting mass profile model for \2315\ requires a central stellar mass component, that the fit improves when the normalisation of this component is left free, and that the addition of adiabatic contraction improves the fit slightly. We emphasise that the radial reach of our observations is sufficient to place strong constraints on the overall form of the mass profile from deep in the central regions ($< 10$ kpc), out to a significant fraction of $R_{500}$. However, the main conclusions  are driven by the fit to the data interior to $\sim 30$ kpc (i.e., the inner four data points), and the effects under discussion (i.e., the normalisation of the stellar mass component and the effect of adiabatic contraction on the dark matter profile) are somewhat degenerate. In the following, we attempt to disentangle the influence of each effect. 

Given that additional stellar mass is clearly needed, the question of the normalisation of the stellar component becomes crucial in interpreting the inner dark matter density slope. While initially the normalisation of the stellar mass component was set to $M_*/L_R = 1.84$ as detailed in Sect.~\ref{sec:nstar}, formally the best fits (both in terms of $\chi^2$ and in terms of dvi$_{\rm max}$) are given by the NFW$^{\ast}\mathrm{AC}+k\,$star and N04$^{\ast}\mathrm{AC}+k\,$star models where the stellar normalisation is left free and there is adiabatic contraction of the dark matter. 


\subsubsection{On the choice of IMF and its impact on the best fitting mass model}
\label{sec:IMF}

As Table~\ref{tab:resfitmass} shows, the NFW and N04 mass profile fits to \2315\ significantly improve when stars are included but no prior is put on the IMF. Fits are better in an absolute sense with adiabatic contraction as implemented in \citeauthor{gne04}'s code; however, if there is no adiabatic contraction, the    $3\sigma$ ranges for $M_{\star}/L_{\mathrm{R}}$ (and thus the IMF) lie between $\sim 1.2$ and $\sim 7.7$ (for the NFW profile) or 0 and $\sim 11.5$ (N04 profile), i.e., it is largely unconstrained. 
This is to be compared with the $\pm 3 \sigma$ confidence range for $M_*/L_R$ derived from the observed $B-R$ color using the large suite of models by \citet{zib09}. The predicted range is of 0.76--2.27 if a \citet{chabrier03} IMF is assumed, while it is 1.33--3.97 for a \citet{sal55} IMF. Thus, with a central value of $M_{\star}/L_{\mathrm{R}} \approx 4$, the models without adiabatic contraction seem to favour more bottom-heavy IMFs like that of \citet{sal55}.

Interestingly, when an NFW profile is assumed and adiabatic contraction is applied to the dark matter, the uncertainty on $M_{\star}/L_{\mathrm{R}}$ drops by a factor of three, and the robust best-fitting value of $M_{\star}/L_{\mathrm{R}}$ is remarkably consistent with the value predicted assuming a \citet{chabrier03} IMF. Similar conclusions can be drawn in the case of an N04 profile, but they are much weaker since the uncertainties on the values of $M_{\star}/L_{\mathrm{R}}$ become larger. In other words, inclusion of adiabatic contraction leads naturally to an $M_{\star}/L_{\mathrm{R}}$ that is consistent with a Chabrier IMF whereas the models without  adiabatic contraction tend to prefer a Salpeter IMF.

Hence, the assumption of a particular dark matter profile seems to impact on the robustness of the conclusions on the IMF. 
Distinguishing between IMFs at the low-mass end is an
extremely challenging task: in fact, although low mass stars
contribute significantly to the mass, their optical/near-IR flux is negligible
and thus photometric measurements at these wavelengths provide very
weak constraints. While the inclusion of more bands (especially
near-IR) would partly restrict the possible range of $M_*/L_R$ {\it at
fixed IMF}, this would not alleviate the systematic uncertainty
deriving from the unknown IMF.

The standard \citet{sal55} IMF corresponds to a single-slope power law
$\phi (M) \propto M^{-s}$ for $0.1 < M < 100~\mathrm{M}_{\sun}$,
where $s = 2.35$. It provides the lowest value of the stellar mass-to-B-band luminosity ratio ($M_{\star}/L_{\mathrm{B}}$) with respect to analogous IMFs
with a very steep (dwarf dominated) or a very flat (remnant dominated) slopes. Yet for nearby elliptical galaxies, suitable simple stellar population (SSP) models with ages of 12 Gyr predict a value of $M_{\star}/L_{\mathrm{B}}$
which is twice as large as that inferred from dynamical models (under the assumption of a constant stellar mass-to-light ratio, e.g., \citealt{vdm91}).
This excludes any single-slope IMF and enforces a flattening of the IMF
with respect to the Salpeter slope below $\sim 0.5$--$0.7~\mathrm{M}_{\sun}$ \citep[e.g.,][]{ren05}. Consistently, direct stellar counts in Galactic globular clusters \citep{par00} and young clusters with ages ranging
from a few Myr to 1 Gyr (\citealt{dem05}; see also \citealt{bou05}) point to an IMF with a log-normal form below $1~\mathrm{M}_{\sun}$. This conclusion holds for the present day mass function of fields in the Galactic Disc \citep{chabrier03,mor04} or Bulge \citep[][and references therein]{zoc05}.

Interestingly, in an analogous {\em Chandra} study of mass profiles in seven elliptical galaxies with either galaxy-scale or group-scale halos \citep{hum06}, the stellar mass-to-K-band luminosity ratio ($M_{\star}/L_{\mathrm{K}}$) was found to be consistent with SSP models assuming a \citet{kro01} IMF. The latter contains a flattening below $\sim 0.5$--$0.7~\mathrm{M}_{\sun}$, similar to a \citet{chabrier03} IMF. 

The IMF is also indicated as the most significant source of systematic uncertainty in a recent test of adiabatic contraction using profiles of 75,086 elliptical galaxies from the Sloan Digital Sky Survey \citep{sch09}. This study is based on weak lensing observations in the outskirts of the halo and measurements of the stellar velocity dispersion in the inner regions of galaxies for stacked galaxy samples. Schulz et al. conclude that stellar masses need to be larger by a factor of two with respect to those obtained with a \citeauthor{kro01} IMF to explain the inner dynamical-mass excess in their data without adiabatic contraction, but such an increase would create tension with results from SAURON \citep{cap06}.

From this discussion, we conclude that fits to the mass profile of RXC\,J2315.7$-$0222 yielding low values of $M_{\star}/L_{\mathrm{R}}$ (i.e., with a Chabrier-like IMF) must be preferred on physical grounds. This implies that some form of adiabatic contraction has to be invoked whatever the underlying dark matter profile. This conclusion is at odds with that of \citet{hum06}, who cast some doubt on the \citet{gne04} adiabatic contraction scenario since their best-fitting NFW*AC$+$star models to early type galaxy mass profiles yielded significantly lower values of $M_{\star}/L_{\mathrm{K}}$ than predicted by a \citet{kro01} IMF. At variance with their best-fits, we do see a significant change in the uncertainties associated with the best-fitting values of $M_{\star}/L_{\mathrm{R}}$ when adiabatic contraction is applied, at no cost of tension with the values predicted by a range of IMFs. However, we do confirm that allowing adiabatic contraction does not produce evident improvements in the significance of the best fits, irrespective of the halo profile. 

The different conclusion regarding the role played by adiabatic contraction may result from the limited spatial extent of the halo and galaxy regions probed in the study of \citet{hum06}; alternatively, the dissimilar nature of the objects under study in the present work and in that of \citeauthor{hum06} (early-type galaxies vs. fossil groups) may cause other conclusions to be drawn.


\subsubsection{Characteristics of the dark matter profile}

As mentioned in the introduction, fossil systems are so named because they are supposed to be the endpoint of the merger history of an early-forming compact group and as such their dark matter profile should reflect their early age of formation in a higher concentration than average. An important question that remains to be addressed is whether the concentration of the dark matter profile of \2315\ is any different from typical literature values for non-fossil systems of a similar mass. 

\begin{figure}[]
\includegraphics[width=8cm]{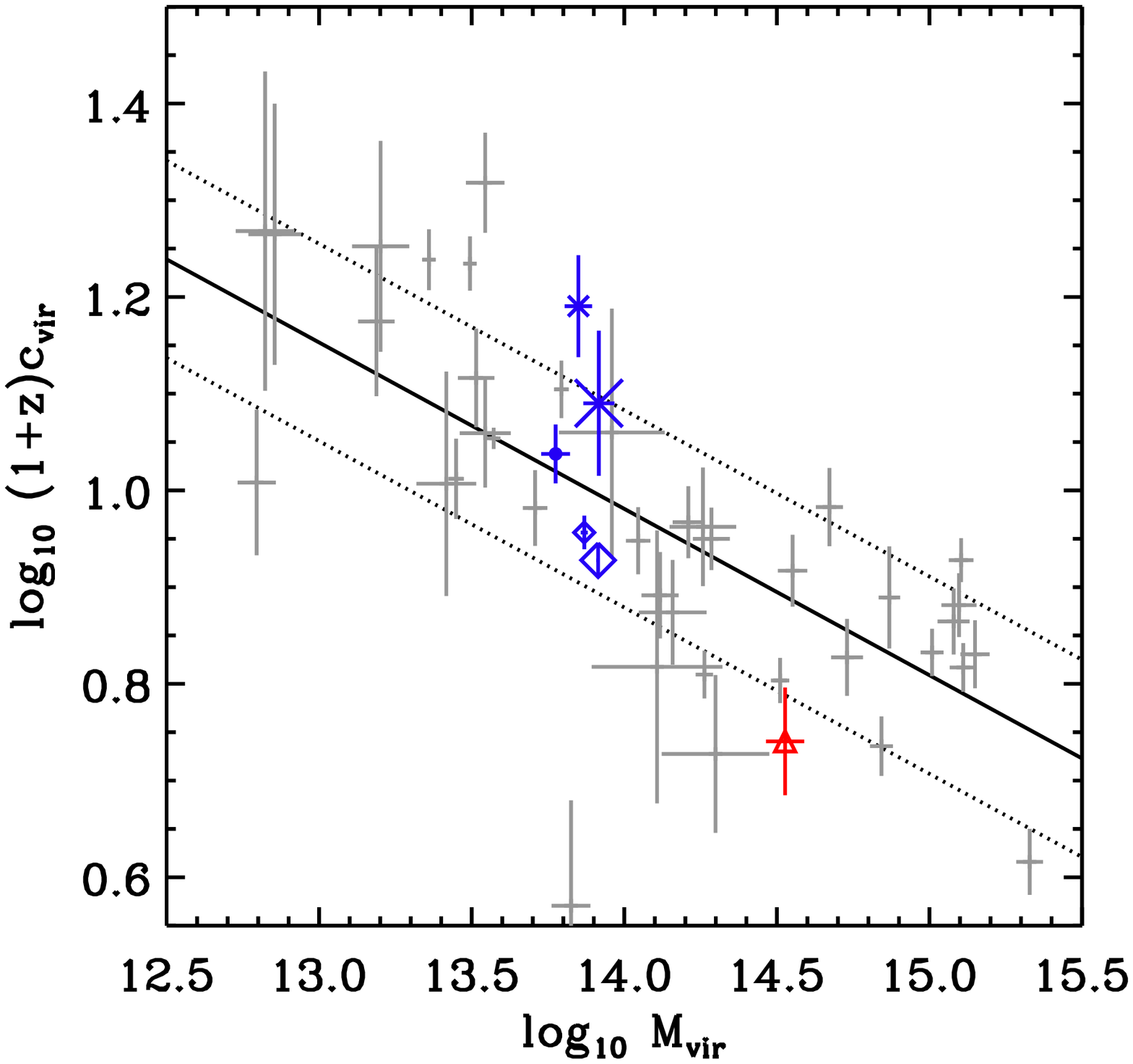} 
\caption{Fossils on the $c_{vir}-M_{vir}$ compilation from \citet{buo07}. Normal cluster and group data are plotted in grey; the black solid line is the best fitting $c_{vir}-M_{vir}$ relation found by \citeauthor{buo07} and the dotted line the $1\sigma$ uncertainties. Open triangle: \0216\ NFW fit with AGN contribution removed. All other symbols refer to \2315. Filled circle: NFW fit. Small asterisk: NFW+star. Large asterisk: NFW+$k\,$star. Small diamond: NFW*AC+star. Large diamond: NFW*AC+$k\,$star.\label{fig:concentration}}
\end{figure}

\begin{figure*}[]
\includegraphics[width=0.48\textwidth]{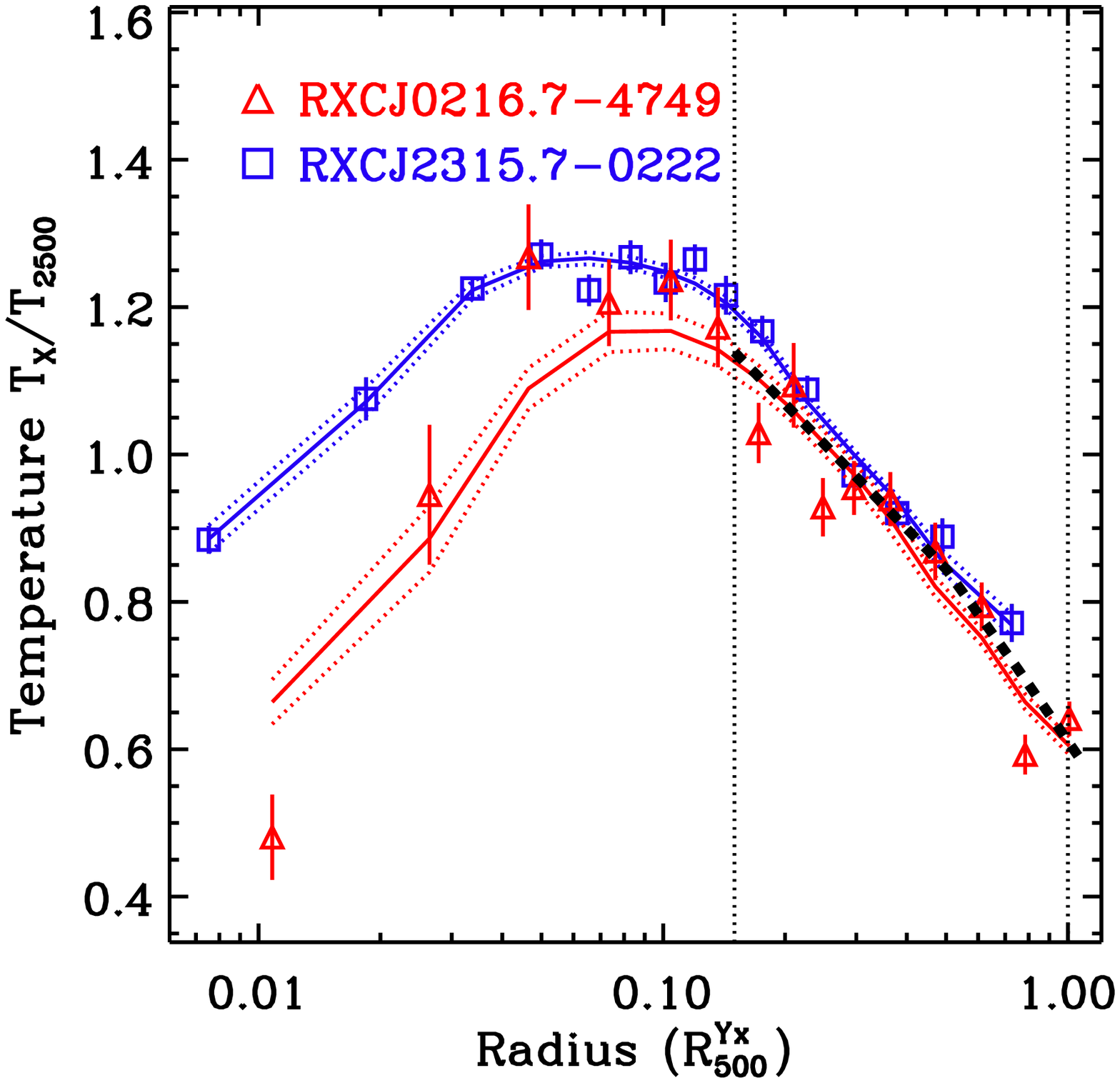} 
\hfill
\includegraphics[width=0.48\textwidth]{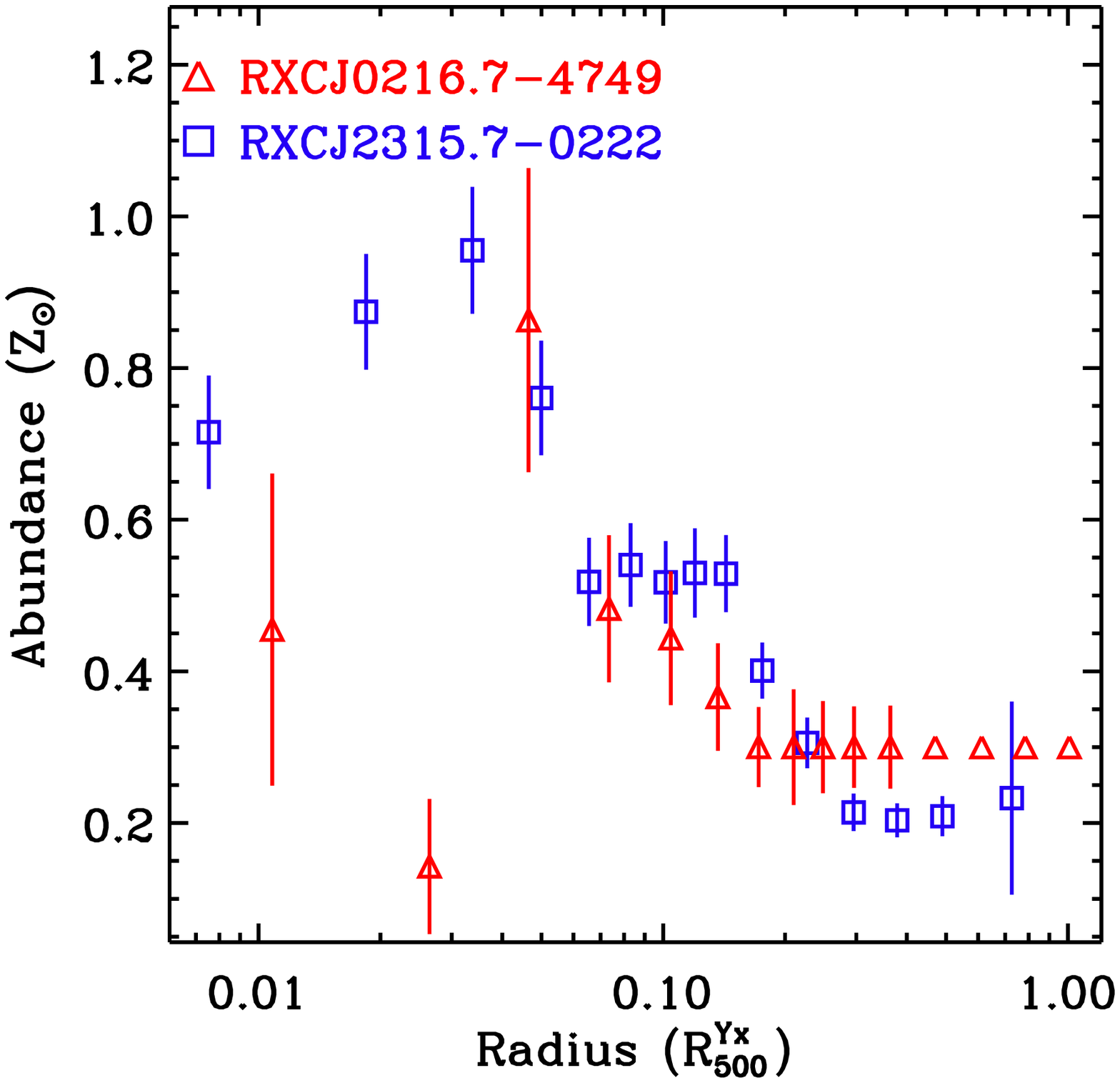} 
\caption{{\it Left panel}: Temperature profiles scaled by $R_{500}^{Y_X}$ (646 kpc and 569 kpc, for \0216\ and \2315, respectively) and $T_{2500}$ (2.25 keV and 1.41 keV for \0216\ and \2315, respectively), where $T_{2500}$ is the spectral temperature in the aperture  $[0.15-1]\,R_{2500}$. The black dashed line shows the best fitting slope in the radial range $[0.15-1]\,R_{500}$. {\it Right panel}: Abundance profiles scaled by $R_{500}^{Y_X}$. Points with no error bars are frozen at $Z/Z_\odot = 0.3$. \label{fig:scprofs}}
\end{figure*}

The simple NFW fit to the total mass density profile leads to the highest concentration of any of the model fits, supporting the conclusions of \citet{mam05}, who suggested that neglect of the central stellar mass in the central regions, together with a restricted radial fitting range, could mimic a high mass density concentration, as was found by some earlier work \citep[e.g.,][]{kho04}\footnote{Note that the extremely high concentration found for \object{NGC 6482} by \citet{kho04} may also have exacerbated by the limited radial range of their data.}. We find that depending on the mass model, the resulting values of $M_{500}$ only cover a relatively small range from $3.6 - 4.4 \times 10^{13}\, M_{\odot}$, with corresponding concentration parameters $c = 5-8$. Applying adiabatic contraction to the dark matter generally lowers the concentration (see Table~\ref{tab:resfitmass}). 

Figure~\ref{fig:concentration} shows the concentration parameter derived from the various NFW model fits compared to the data compilation of relaxed systems from \citet{buo07}\footnote{We compute $\Delta_{vir}$ using $\Delta_c(z) = 18\pi^2 + 82\times [\Omega_m (z) - 1]-39\times[\Omega_m (z) - 1]^2$ with $\Omega_m (z) \equiv \Omega_m (z = 0) = 0.3$ \citep[see][]{bry98}.}. The two fossil groups in the present analysis do not exhibit particularly high or low concentrations in view of the considerable dispersion in the range of measured values. Furthermore, the small span of values we find for both $c$ and $M_{500}$ suggest that we have sufficient radial leverage to make an unbiased estimate of these parameters. 


\subsubsection{Effect of possible central point source}\label{sec:psource}

The mass profile results discussed above for \2315\ were derived assuming that there is no additional central point source that emits in X-rays. As this might actually be the case, we investigated the possibility that a second component may exist by fitting the spectrum of the inner three annuli with a power law of fixed slope $\Gamma = 1.4$ in addition to the thermal emission model. In the innermost spectral region, a better fit is obtained with the MeKaL plus powerlaw model (confirmed by an F-test), suggesting that a central X-ray source may indeed be present and contributing $\leq 10$ per cent of the emission in that region\footnote{The upper limit to the X-ray liminosity of this source in the 2-10 keV band is $L \sim 2.9\times 10^{39}$ erg s$^{-1}$ which is consistent with the luminosity expected from low mass X-ray binaries according to \citet{grimm2002} study of X-ray binaries in the Galaxy.}. We thus corrected for the effect of this central source on the density and temperature profiles as described in Appendix~\ref{app:0216AGN} and refitted the mass profile with the mass models described above in Section~\ref{sec:massmods}. The best fitting model parameters are given in Table~\ref{tab:resfitmasscen}.

While there is a slight change in concentration towards lower values (as expected), the most important result is that the same trends and conclusions are valid for this analysis as for the analysis assuming no central source. We are thus confident that our conclusions regarding the properties of the mass and dark matter profiles of this system are robust to the presence of a central X-ray point source (if any).


\begin{figure*}
\includegraphics[width=0.32\textwidth]{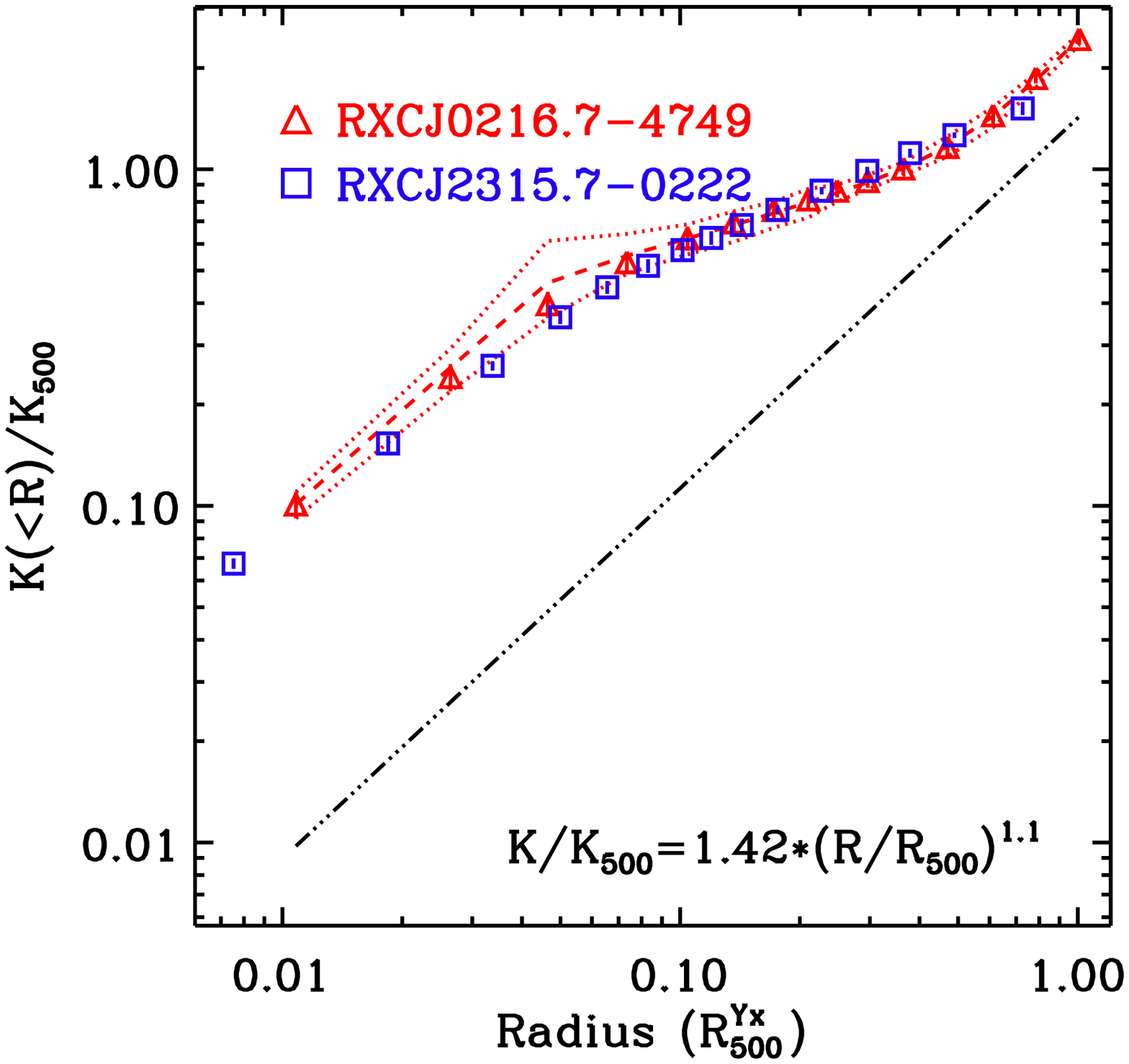} 
\hfill
\includegraphics[width=0.32\textwidth]{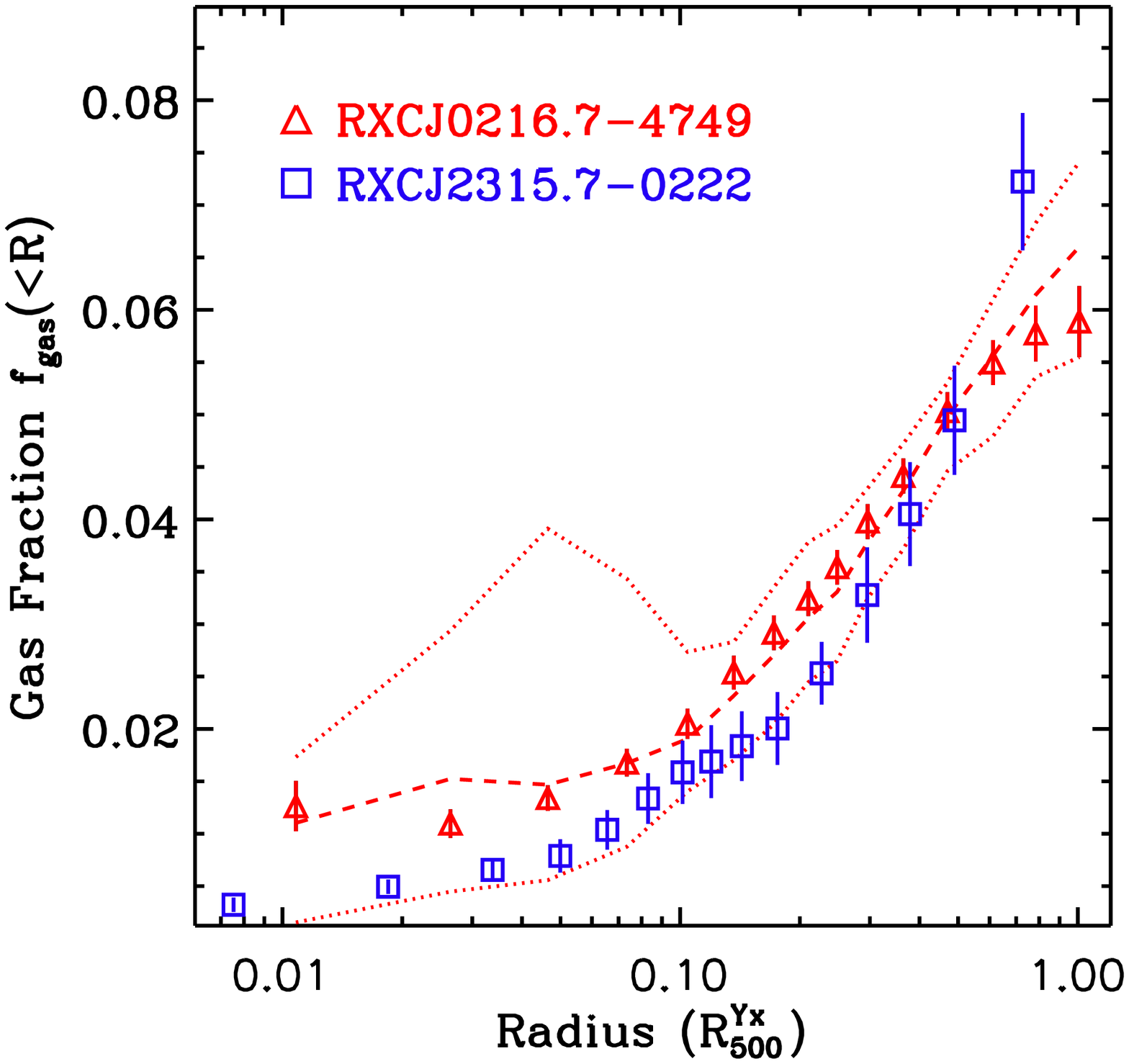} 
\hfill
\includegraphics[width=0.32\textwidth]{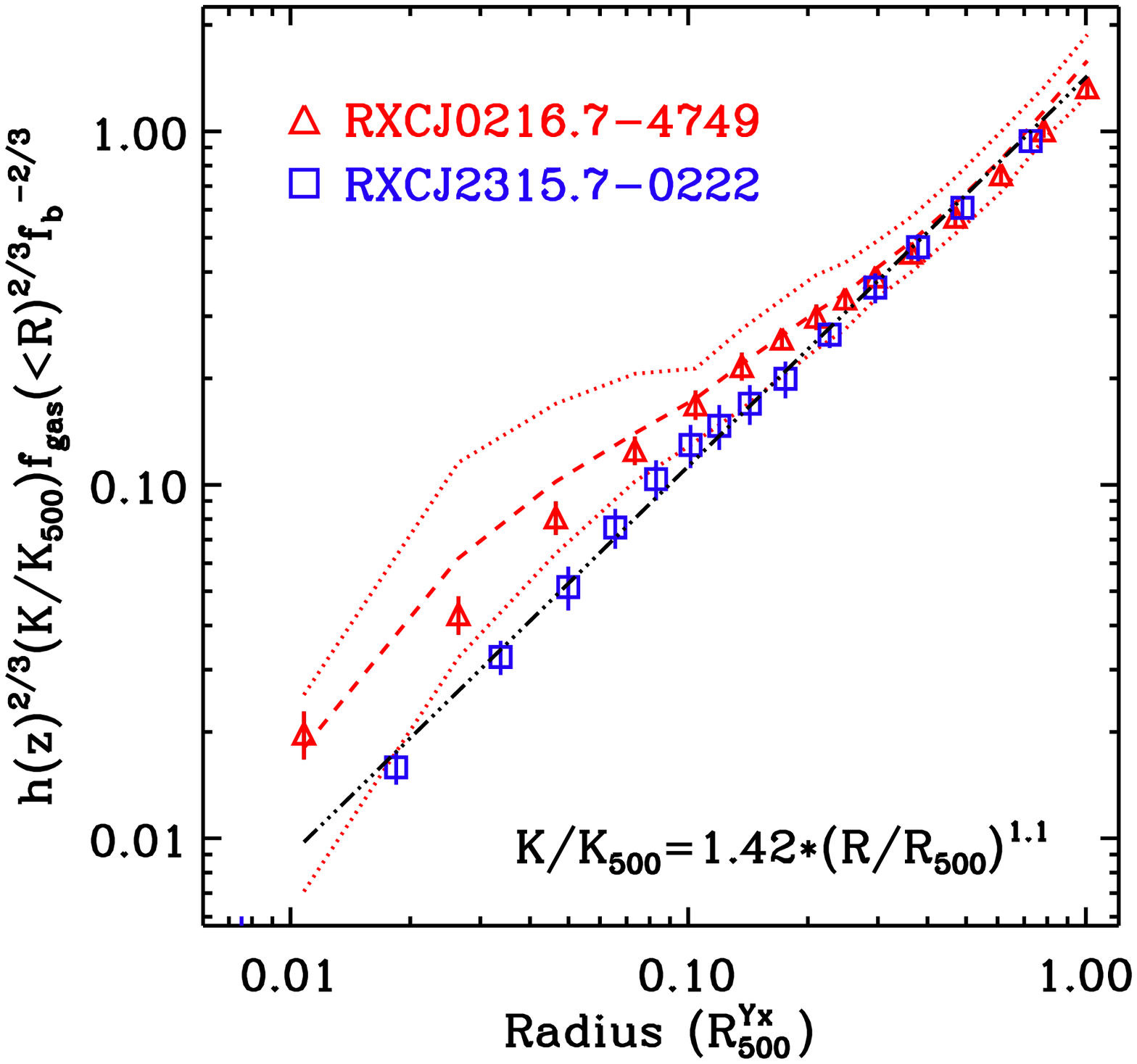}
\caption{{\it Left panel}: Entropy scaled by $R_{500}^{Y_X}$ and $K_{500}$. {\it Middle panel}: Gas mass fraction profiles. {\it Right panel}: Scaled entropy multiplied by gas mass fraction. In left and right hand panels, the dashed line shows the baseline entropy prediction of \citet{vkb05}. \label{fig:scaledK500}}
\end{figure*}

\subsection{Gas properties}

\subsubsection{Scaled temperature and abundance profiles}

The scaled temperature and abundance profiles of the two fossil groups are shown in Figure~\ref{fig:scprofs}. When plotted with a logarithmic radial axis, the temperature profiles for both of the groups exhibit the bell shape typical of cool core clusters \citep[e.g.,][]{vikh06,pratt07}. The temperature peak is found interior to $0.1\,R_{500}$ in each case, further in than is observed in clusters and in agreement with the findings of \citet{sun09} from the analysis of a large sample of groups. Outside the core regions the temperature profiles both decline with a similar slope. We fitted the combined data set in the $[0.15-1]\,R_{500}$ radial range with the model $T/T_{2500} = N \times (1+ r/R_{500})^{\alpha}$, for which we find $N = 1.33\pm 0.03$ and $\alpha = -1.13\pm 0.05$. These values are consistent with those found by \citet{sun09}, suggesting that fossils are not exceptional in the properties of their inner or outer temperature profiles when compared to other systems. The abundance profiles of the two groups are substantially similar, exhibiting an off-centre abundance peak similar to that seen in some other observations of group scale objects \citep[e.g.,][]{rp07}, where the mean profile rises towards the central regions, then exhibits a drop in the inner $\sim20$ kpc.


\subsubsection{Entropy and gas mass fraction}

The entropy profiles of the two groups are shown in the left hand panel of Figure~\ref{fig:scaledK500}. As is conventional, `entropy' is calculated from the density and temperature profiles: $K = kT/n_e^{2/3}$; in addition, they have been scaled by the characteristic entropy of the system

{\small 
\begin{equation}
K_{500} = 106 \ {\rm keV\ cm}^{-2} \left( \frac{M_{500}}{10^{14}\,h_{70}^{-1}\,M_\odot} \right)^{2/3}\, \left(\frac{1}{f_b}\right)^{2/3}\, E(z)^{-2/3}\, h_{70}^{-4/3} \label{eqn:K500}
\end{equation}
}
\noindent \citep[e.g.,][]{vkb05,pra09}. The dashed line shows the best fitting power law fit to the median entropy profile in the radial range $[0.1 - 1]\,R_{200}$ for the clusters formed in the non-radiative simulations of \citet{vkb05}. The observed profiles exhibit much the same form, and lie very significantly above the baseline prediction, as is expected if non-gravitational processes affect the ICM to a greater extent at the group scale. The entropy excess is significant even at large radius (and in the case of \0216, this is true at $R_{500}$), in agreement with the results of \citet{sun09} and in contrast to the relative lack of excess at large radius found in clusters \citep{nkv07,sun09,pra09}. The radial entropy slope is shallower than the typical value of 1.1 except in the very outer regions ($r \gtrsim 0.6\,R_{500}$).

The middle panel of Figure~\ref{fig:scaledK500} shows the integrated gas mass fraction profiles $f_{\rm gas} (< R) = M_{\rm gas} (< R) / M (<R)$. The gas mass fraction increases with radius in each case. The average gas mass fraction at $R_{2500}$ is $f_{\rm gas, 2500} = 0.045\pm0.005$, consistent with the results of the larger sample of groups studied by \citet{sun09}. As Table~\ref{tab:resfitmass} shows, measurement of the gas mass fraction of \2315\ at $R_{500}$ is somewhat model dependent. The simple NFW model, which gives the worst fit to the overall mass profile, yields the highest value of $f_{\rm gas,500}$ due to the model's systematic underestimate of the data point at $\sim 400$ kpc (Figure~\ref{fig:2315mprofs}). The most reliable values can be obtained from the best fitting models, yielding $f_{\rm gas,500} \sim 0.087$ for this group, a value consistent with those found by \citet{sun09} for systems of similar temperature.

The link between increased entropy and the total gas content was recently demonstrated by \citet{pra09}, who showed that multiplication of the scaled entropy profile with the scaled gas mass fraction profile, in effect correcting the entropy for the difference in total gas content with radius, yielded entropy distributions that were in good agreement with the predictions from adiabatic simulations. We show the corresponding fossil group profiles in the right hand panel of Figure~\ref{fig:scaledK500}; clearly, this correction also works on the group scale, providing further evidence that gas content is the key to understanding the physical processes responsible for the behaviour of the entropy.


\section{Conclusions}

We have presented the first deep X-ray and wide field optical imaging observations of two candidate fossil groups, \0216\ and \2315. Based on the criteria established by \citet{jon03}, and taking into account the various uncertainties involved in the definition of $R_{500}$, we argue that \0216\ is a  {\it bona fide} fossil system and \2315, if not formally fossil, shares strong physical similarities with this type of object. The X-ray data quality is exceptional for this type of object, extending from $[0.01 - 0.75]\,R_{500}$ in both cases, allowing us to investigate in detail the properties of their profiles. While unfortunately the central regions of \0216\ are contaminated by a bright X-ray point source that contributes $\sim 40$ per cent of the emission in the central temperature profile bin, we devise a method to correct for its presence and derive the resulting corrected density and temperature profiles. 

The object temperatures are $2.05\pm0.05$ keV and $1.68\pm0.03$ keV for \0216\ and \2315, respectively, when measured in the $[0.15-1]\,R_{500}$ region, placing them squarely in the galaxy group category. Both systems exhibit regular, highly peaked X-ray emission centred on the BCG, indicative of their being morphologically relaxed objects. The temperature profiles both describe the shape typical of cool core systems, but with a temperature peak at $\sim 0.075\,R_{500}$, closer to the centre than is observed for more massive systems. Their entropy profiles show a considerable excess above the expectations from non-radiative simulations across the entire measured radial range (this is true out to $R_{500}$ for \0216). 

Using the temperature and density profiles, and assuming hydrostatic equilibrium, we calculated the total mass profiles of the two groups. For \0216, the best fitting NFW model yields $c_{500} = 2.22\pm0.25$ and $M_{500} = 1.31^{+0.12}_{-0.11} \times 10^{14}\ M_\odot$; however, mass constraints overall are weak due to the uncertainties associated with correction for the AGN, and we do not fit more complex models to these data.

The mass profile of \2315\ is of sufficient quality for deeper investigation. We find that consideration of the stellar mass of the central galaxy is essential to provide a good fit to the data. The best fitting mass model is either the S\'ersic-like profile proposed by \citet{nav04} with an index $\alpha$ in agreement with predictions, or an NFW profile, in each case plus a stellar component. The concentration is not especially high compared to non-fossil systems, and appears to be in the range observed and expected for normal systems. Applying adiabatic contraction to the dark matter improves the fit slightly and consistently yields a lower $M_*/L_R$ ratio. Based on the range of derived $M_*/L_R$ ratios and comparison to a range of literature IMFs, we argue that low $M_*/L_R$ fits are preferred on physical grounds, implying that adiabatic contraction has operated in this system. These conclusions are robust to the presence of a possible central X-ray point source.

Clearly the most significant source of uncertainty in our analysis is the IMF. Observation of more extreme fossil systems may allow unambiguous detection of the adiabatic contraction effect on the dark matter, and provide evidence that these systems are indeed older than normal systems. As it stands, in the presence of such excellent data, better observational and theoretical progress on the IMF is necessary to draw definitive conclusions.


\begin{acknowledgements}
We thank R. Piffaretti for useful discussions, J. Thomas for help with deprojection of the stellar luminosity profile, H. B\"ohringer for help with the initial target selection, and the referee for a useful report. The present work is based on observations obtained with {\it XMM-Newton}, an ESA science mission with instruments and contributions directly funded by ESA Member States and the USA (NASA). IRAF is the Image Reduction and Analysis Facility, a general purpose software system for the reduction and analysis of astronomical data. IRAF is written and supported by the IRAF programming group at the National Optical Astronomy Observatories (NOAO) in Tucson, Arizona. NOAO is operated by the Association of Universities for Research in Astronomy (AURA), Inc. under cooperative agreement with the National Science Foundation.
\end{acknowledgements}

\bibliographystyle{aa} 
\bibliography{fgroups}


\appendix

\section{Background processing for \2315 \label{app:2315background}}   

In the case of \2315\, the source emission fills the {\it XMM-Newton} field of view, limiting the application of our standard method, which relies on a source-free region from which to estimate the local background, and ultimately, the CXB  contribution to the annular cluster spectra. For this system, we thus adopted the following procedure to calculate the surface brightness and temperature profiles.

\subsection{Surface brightness profile}

We take the surface brightness profile subtracted from the FWC background data and fit it with the analytical model proposed by \citet{vikh06} plus a constant to estimate the CXB background level. We then take the mean between the CXB estimated with our standard method (which is an overestimate) and the constant level estimated from the fit. We then checked our estimate of the CXB contribution using {\it ROSAT\/} data from a region around the position of \2315, finding that the estimates are consistent. We use the mean CXB estimate and add the difference between the result from the standard method and the analytical model plus constant model fit as a systematic error.

\subsection{Spectral analysis}

An estimate of the spectral contribution from the CXB was obtained by simultaneously fitting the two outermost annuli with a model consisting of two unabsorbed thermal components corresponding to the galaxy and the local bubble, an absorbed powerlaw with slope fixed to $\Gamma = 1.4$ corresponding to the contribution from unresolved AGN, and an absorbed thermal emission model representing the group component. Physical background parameters were linked between the two areas (temperatures, powerlaw index, etc) but the normalisation of the background contribution was allowed to vary. The final best fitting background model is then used in the fit to the inner annuli, with its normalisation scaled to the ratio of the areas under consideration.

We checked that the CXB contribution estimated from the spectral analysis was consistent with that obtained from the surface brightness profile.


\section{Estimating the AGN contribution for \0216 \label{app:0216AGN}}   

\begin{figure*}[ht]
\includegraphics[width=0.32\textwidth]{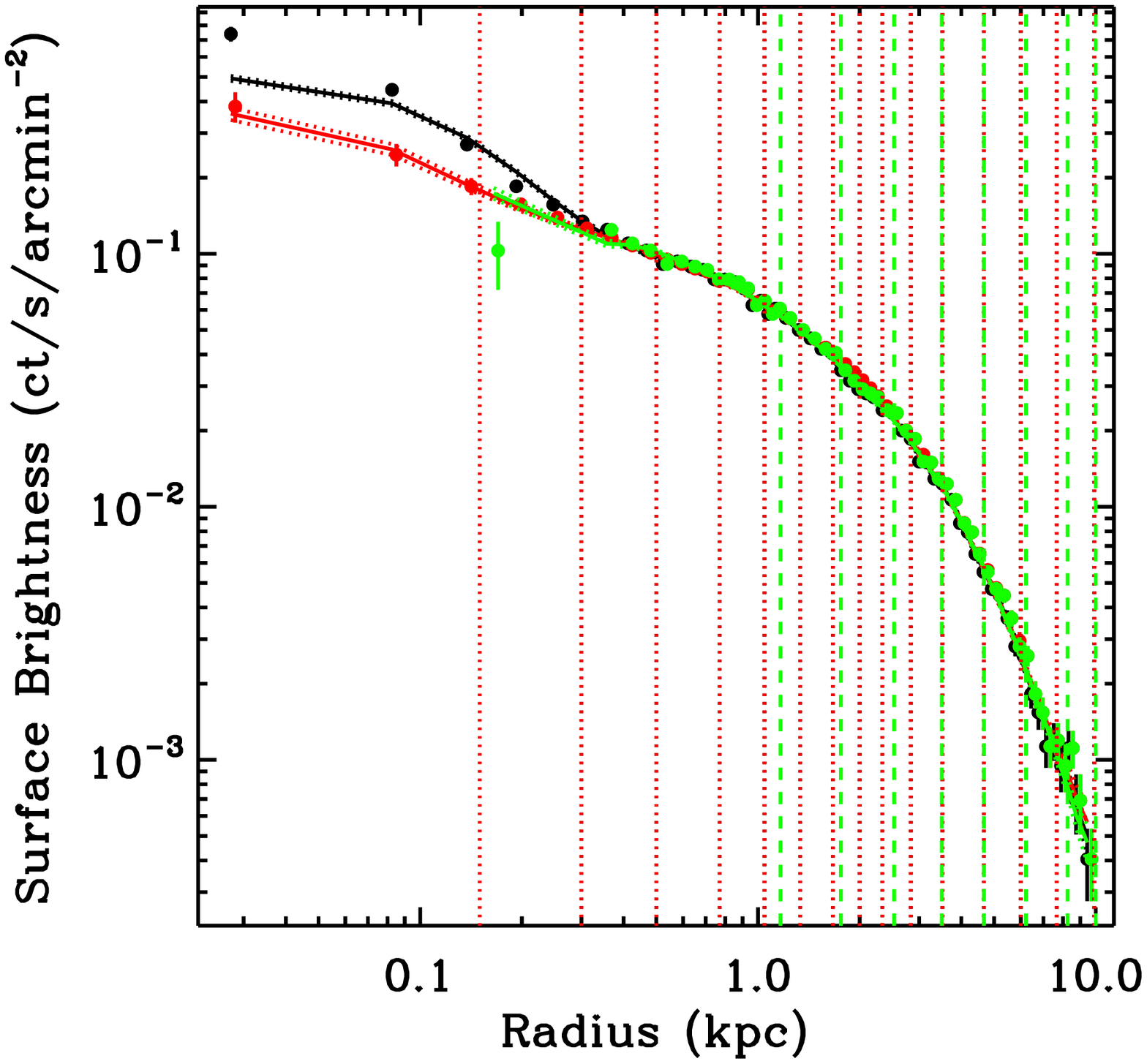} 
\hfill
\includegraphics[width=0.32\textwidth]{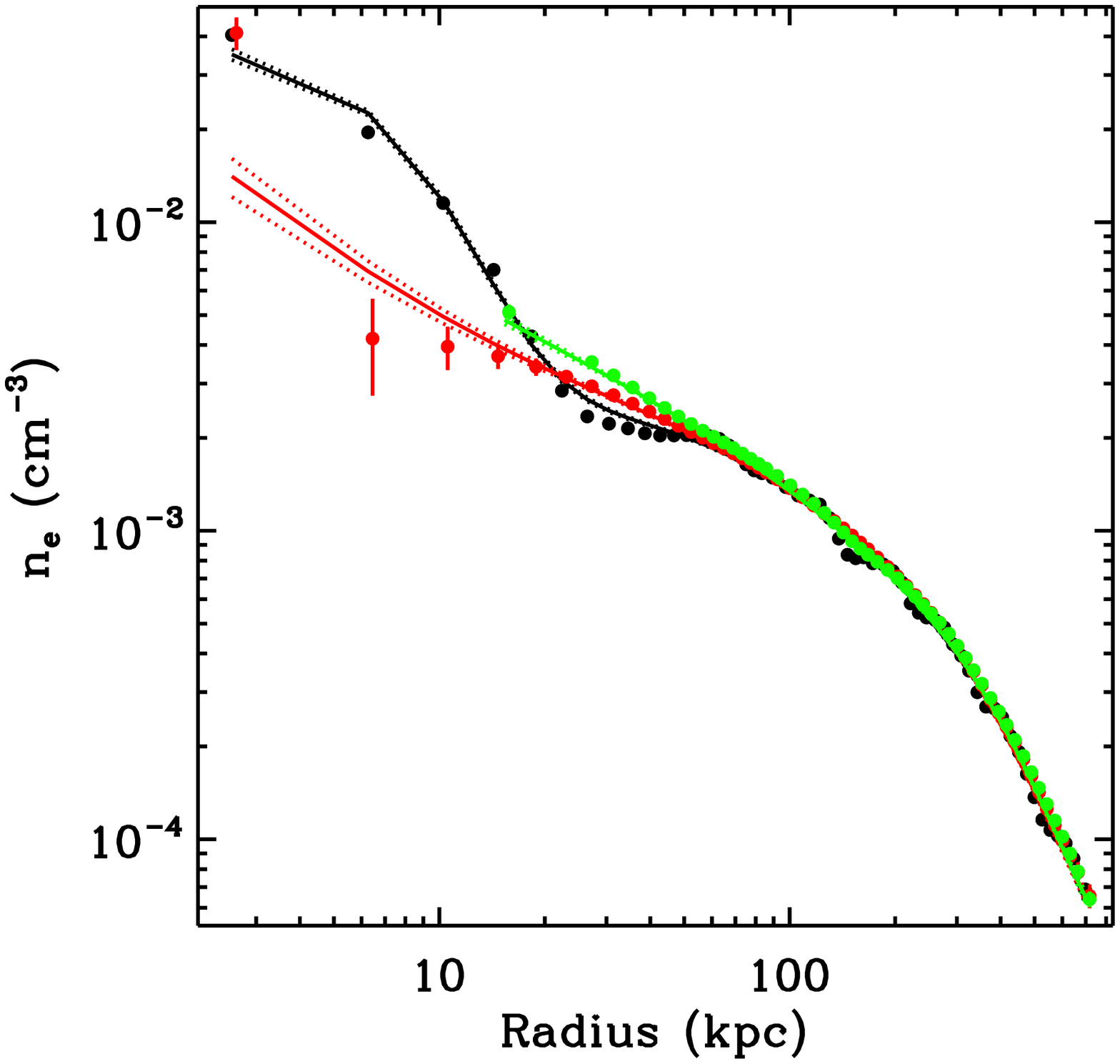}
\hfill
\includegraphics[width=0.32\textwidth]{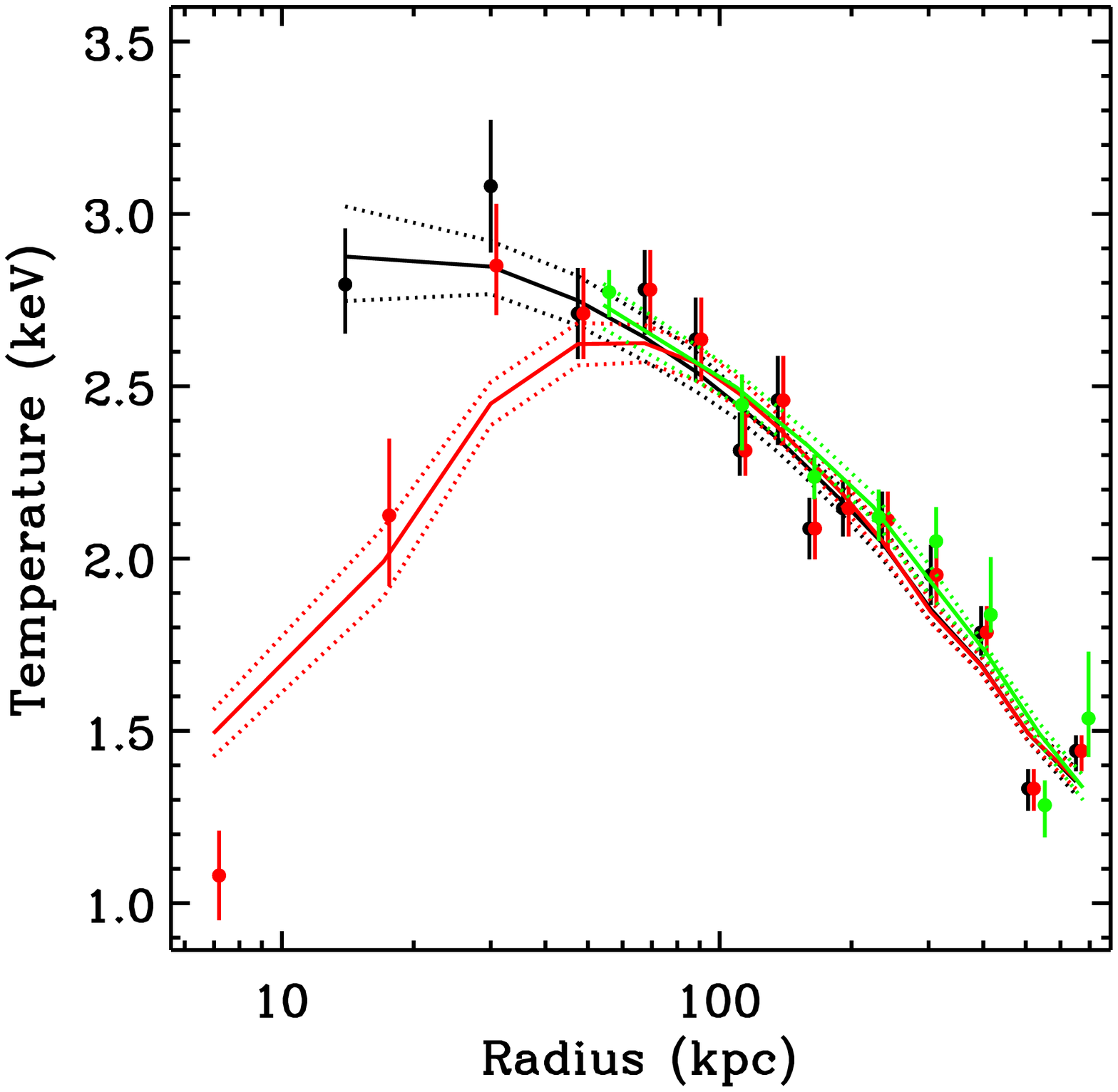} 
\caption{Surface brightness (left), density (middle) and temperature (right) profles of \0216. Black: standard analysis with no correction for the AGN contribution (the inner point of the temperature profile is omitted because the single thermal model gives a $\chi^2/\nu > 2$).  Green analysis after excision of the inner $18\arcsec$ of profile data. Red: analysis with AGN contribution corrected as described in this Appendix.}\label{fig:0216sumprof}
\end{figure*}

\0216 was found to have an X-ray bright AGN at its centre, manifesting itself after our standard deconvolution in  a very cusped density profile and a flat inner temperature profile (see Fig.~\ref{fig:0216sumprof}). Various tests confirmed our suspicion; for instance  an F-test showed that the addition of a power law component in the central regions considerably improved the spectral fit, and the surface brightness profile was better fitted with an AB model \citep{pa02} plus a point source convolved with the PSF.

We first sampled the central region in such a way as to maximise the AGN contribution to the innermost temperature profile bin. In this $9\arcsec$ region, a single thermal emission model fit yields a $\chi^2/$dof$\sim 3$ and a  temperature of $kT\sim 2.5$keV. Adding an absorbed powerlaw improves the fit considerably (confirmed with an F-test), and suggests that the AGN contributes 40-50 per cent of the total counts in this region. We used several different methods to estimate the AGN contribution, as listed below.

\begin{enumerate}
\item Fitting the surface brightness profile with an analytical model \citep{vikh06} plus a point source:

\begin{itemize}
\item fit the surface brightness profile with the model (fossil plus point source);
\item integrate each component of the result (fossil and AGN) in each annulus;
\item fit the spectra of the annuli and tune the powerlaw normalisation so as to find a consistent count rate for the powerlaw component in the 0.3-2 keV band (corresponding to the extraction band of the surface brightness profile).
\end{itemize}

\item Fitting the surface brightness profile with an AB profile model \citep{pa02} plus point source, repeating the steps in 1 above. This alternative  was introduced because the analytical model of \citet{vikh06} is designed for cool core systems, whose brightness distributions behave differently to a point source when convolved with a PSF.

\item Fitting the spectrum of the central region with a thermal emission model plus an absorbed powerlaw:

\begin{itemize}
\item fit the spectrum of the first annulus with a source plus powerlaw model. The free parameters are the fossil properties (temperature, abundance,...) and the normalisation of the powerlaw component. The powerlaw index is fixed to $\Gamma = 1.4$. We use as a contraint the fit result of the first annulus only because the AGN contribution decreases rapidly outside this region and the spectral fit tends to find zero contribution at odds with the expectation from the surface brightness analysis;
\item estimate the AGN count rate in the 0.3-2 keV band from the spectral fit;
\item fit the surface brightness profile with an analytical model \citep{pa02}plus point source, with the point source normalisation tuned to obtain an AGN contribution consistent with the results from the spectral fit.
\end{itemize}
\end{enumerate}

We consider the final method to give the best estimate of the AGN count rate contribution as it combines both imaging and spectroscopic constraints. Using this method the AGN contribution in the first three temperature profile annuli of \0216\ is estimated to be 40, 18 and 6 per cent, respectively. To compute the uncertainty on the AGN contribution we change its normalisation in XSPEC so as to obtain $\Delta \chi^2 \sim 1.$ fixing all the other parameters. The resulting relative error is $\sim 3.5$ per cent.


\begin{figure}
   \includegraphics[width=8cm]{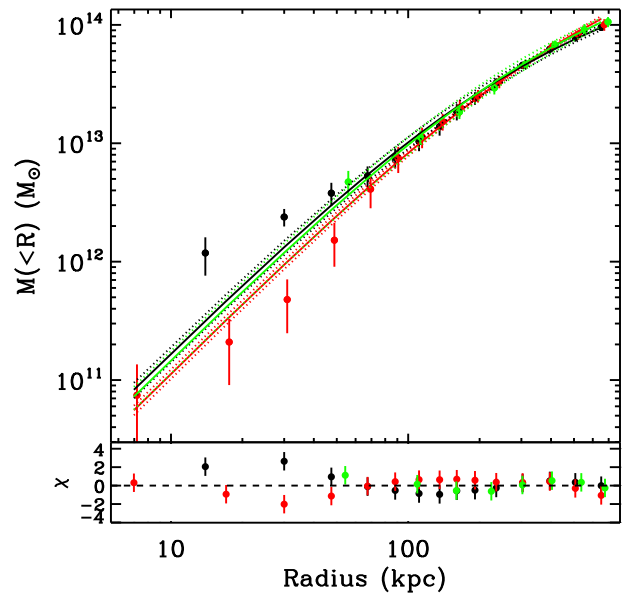} 
  \caption{The mass profile of \0216. Data points are the mass profile derived from hydrostatic equilibrium and solid lines are the best fitting model in each case. Black: standard analysis with no correction for the AGN contribution (the inner point is omitted because the single thermal model gives a $\chi^2/\nu > 2$).  Green: analysis after excision of the inner $18\arcsec$ of mass profile data. Red: analysis with AGN contribution corrected as described in this Appendix.}\label{fig:0216sumMassprof}
\end{figure}

\section{Effect of central regions on mass profile modelling of \0216 \label{app:0216mprof}}

In this Appendix, we are interested in the effect of various ways of dealing with the AGN on the parameters resulting from an NFW model fit to the mass profile. The three cases we considered are illustrated in Figure~\ref{fig:0216sumprof}, which shows the surface brightness (left panel), deconvolved density (middle panel) and deconvolved temperature (right panel) profile of each case. Figure~\ref{fig:0216sumMassprof} shows the resulting mass profiles and best fitting NFW models; model parameters are listed in Table~\ref{tab:resfitR0216}. The three cases are:

\begin{itemize}

\item No correction (black profiles). Here the point source is clearly visible in the surface brightness profile, and our standard deconvolution process yields a strongly peaked density profile. The temperature profile is flat in the inner regions. The resulting mass profile has an excess of mass in the centre compared to the NFW model fit, which does not fit the data points very well ($\chi^2$/dof = 15.0/12). This fit gives the lowest value of $M_{500}$ and the highest concentration.

 \item Exclusion of the central $18\arcsec$ (green profiles). In this case we lose all constraints on the core mass profile of the group, relying on the outer data points to constrain the shape of the mass profile. The NFW model fit is good ($\chi^2$/dof = 2.4/6), and yields values of $M_{500}$ and $c$ that are intermediate between a fit of the uncorrected profile and a fit of the profile with correction for AGN contamination.
  
 \item Modelling of the AGN contribution as described above in Appendix~\ref{app:0216AGN} (red profiles). Here we have the tightest constraints on the mass profile shape, from a profile that extends from $[0.01 - 1]\,R_{500}$. The NFW fit is again good ($\chi^2$/dof = 10.0/13), but yields the highest value of $M_{500}$ and the lowest value of $c$.

\end{itemize}

\begin{table}
\begin{center}
\begin{tabular}{lccc}
\hline
\hline
NFW fit parameter & TOT & CUTCENTRE & AGNMOD \\
\hline
$M_{500}$ ($10^{14}M\odot$) & $1.0^{+0.07}_{-0.065}$ & $1.16^{+0.13}_{-0.12}$ & $1.3^{+0.12}_{-0.11}$ \\

$c_{500}$ & $3.51^{+0.39}_{-0.39}$ & $2.93^{+0.50}_{-0.50}$ & $2.22^{+0.25}_{-0.25}$ \\
\hline
\end{tabular}
\end{center}
\caption{Effect of excising the central regions (CUTCENTRE) or modelling the AGN contribution to the central regions (AGNMOD) on the parameters derived from an NFW model fit to the mass profile of \0216.\label{tab:resfitR0216}}
\end{table}

\section{Mass modelling results for \2315\ with correction for additional central point source\label{app:2316censource}}

As discussed in Section~\ref{sec:psource}, a fit to the spectrum of the central region suggests the possible presence of a point source contribution at the $< 10$ per cent level. The density and temperature profiles were thus corrected for the presence of this possible central point source as described in Appendix~\ref{app:0216AGN}. The resulting mass profiles were refitted with the mass models described in Section~\ref{sec:massmods}. The best fitting model parameters are given in Table~\ref{tab:resfitmasscen}.

\begin{table*}
{\tiny 
\begin{center}
\caption{Results of fits to the mass profile of \2315 with correction for a possible central point point source.\label{tab:resfitmasscen}}
\begin{tabular}{l c c c c c r r r r}
\hline
\hline
Model & $c_{500}$ & $\alpha$ & $M_{500}$ $(10^{13}M_{\odot})$ & $R_{500}$ (kpc) & $f_{gas,500}$ & $M_{\star}/L_R$ & $M_{500}/M_{500}^{Yx}$ & $\chi^2/dof$ & dvi$_{\rm max}$ \\
\hline
\\
NFW                            & $7.14\pm0.60$ & \ldots & $3.80\pm0.25$ & $509\pm10$ & $0.089\pm0.011$ & \ldots & $0.72\pm0.07$ & $18.3/13$  & $0.55$ \\

NFW$+$star                     & $6.63\pm0.61$ & \ldots & $4.18\pm0.13$ & $523\pm5$ & $0.086\pm0.008$ & 1.84 & $0.78\pm0.06$ & $9.7/13$ & $0.28$\\

NFW$+\,k\,$star             & $5.71\pm0.81$ & \ldots & $4.18\pm0.13$ & $524\pm5$ & $0.086\pm0.008$ & $3.34\pm1.04$ & $0.78\pm0.06$ & $7.7/12$ & $0.25$\\

NFW$^{\ast}$AC$+$star          & $5.31\pm0.29$ & \ldots & $4.18\pm0.13$ & $523\pm5$ & $0.086\pm0.008$ & 1.84 & $0.78\pm0.06$ & $9.29/13$ & $0.36$\\

NFW$^{\ast}$AC$+\, k\,$star & $5.70\pm0.14$ & \ldots & $4.18\pm0.12$ &$524\pm5$ & $0.086\pm0.008$ & $1.38\pm0.31$ & $0.78\pm0.06$ & $7.34/12$ & $0.25$\\
\\

N04                            & $5.27\pm0.84$ & $0.118\pm0.021$ & $4.40\pm0.31$ & $532\pm13$ & $0.084\pm0.011$ & \ldots & $0.82\pm0.08$ & 6.48/12 & $0.18$\\

N04+star                       & $5.43\pm0.92$ & $0.161\pm0.037$ & $4.39\pm0.23$ & $532\pm10$ & $0.084\pm0.010$ & 1.84 & $0.82\pm0.07$ & 6.57/12 &$0.20$\\

N04+$k\,$star               & $5.60\pm1.07$ & $0.177\pm0.092$ & $4.35\pm0.22$ & $530\pm9$ & $0.084\pm0.009$ & $1.58\pm1.97$ & $0.81\pm0.07$ & 6.51/11 & $0.21$\\
N04*AC+star                    & $4.88\pm0.66$ & $0.262\pm0.063$ & $4.25\pm0.21$ & $526\pm9$ & $0.085\pm0.009$ & 1.84 & $0.79\pm0.07$ & 7.31/12 & $0.30$\\

N04*AC+$k\,$star            & $5.42\pm1.14$ & $0.188\pm0.104$ & $4.38\pm0.19$ & $532\pm8$ & $0.084\pm0.009$ & $0.80\pm0.94$ & $0.82\pm0.07$ & 6.36/11 &$0.21$ \\

\\
\hline

\end{tabular}
\end{center}
}
\end{table*}

\end{document}